\documentclass[twocolumn]{aastex701}

\newcommand{\pah}{$3.3~\mu\mathrm{m}~\mathrm{PAH}$~}
\newcommand{\hii}{H\,{\footnotesize\fontfamily{cmr}\selectfont II}~} % HII regions
\newcommand{\erg}{$\mathrm{erg}~\mathrm{s}^{-1}\mathrm{cm}^{-2}\AA^{-1}$}
\begin{document}

%\title{\textcolor{teal}{Unveiling the spectral properties of emerging young star clusters in NGC 628: FEAST NIRSpec/MOS survey design and science proof of concept}}

\title{FEAST: a NIRSpec/MOS survey of emerging young star clusters in NGC 628}

\author[orcid=0000-0002-2199-0977,sname='Faustino Vieira']{Helena Faustino Vieira}
\affiliation{Department of Astronomy, Oskar Klein Centre, Stockholm University, AlbaNova University Centre, SE-106 91 Stockholm, Sweden}
\email[show]{helena.faustinovieira@astro.su.se}  

\author[0000-0002-8192-8091]{Angela Adamo}
\affiliation{Department of Astronomy, Oskar Klein Centre, Stockholm University, AlbaNova University Centre, SE-106 91 Stockholm, Sweden}
\email{angela.adamo@astro.su.se}

\author[0000-0003-1024-7739]{Neville Shane}
\affiliation{Space Telescope Science Institute, 3700 San Martin Drive, Baltimore, MD 21218, USA}
\email{nshane@stsci.edu}

\author[0000-0002-0806-168X]{Linda J. Smith}
\affiliation{Space Telescope Science Institute, 3700 San Martin Drive, Baltimore, MD 21218, USA}
\email{lsmith@stsci.edu}

\author[0000-0001-8068-0891]{Arjan Bik}
\affiliation{Department of Astronomy, Oskar Klein Centre, Stockholm University, AlbaNova University Centre, SE-106 91 Stockholm, Sweden}
\email{arjan.bik@astro.su.se}

\author[0000-0001-8490-6632]{Thomas S.-Y. Lai}
\affiliation{IPAC, California Institute of Technology, 1200 E. California Blvd., Pasadena, CA 91125, USA}
\email{ThomasLai.astro@gmail.com}

\author[0000-0002-8222-8986]{Alex Pedrini}
\affiliation{Department of Astronomy, Oskar Klein Centre, Stockholm University, AlbaNova University Centre, SE-106 91 Stockholm, Sweden}
\email{alex.pedrini@astro.su.se}

\author[0000-0001-9162-2371]{Leslie K. Hunt}
\affiliation{INAF -- Osservatorio Astrofisico di Arcetri, Largo E. Fermi 5, 50125 Firenze, Italy}
\email{leslie.hunt@inaf.it}

\author[0000-0002-1000-6081]{Sean T. Linden}
\affiliation{Steward Observatory, University of Arizona, 933 N Cherry Avenue, Tucson, AZ 85721, USA}
\email{seanlinden@arizona.edu}

\author[0009-0003-6182-8928]{Giacomo Bortolini}
\affiliation{Department of Astronomy, Oskar Klein Centre, Stockholm University, AlbaNova University Centre, SE-106 91 Stockholm, Sweden}
\email{giacomo.bortolini@astro.su.se}

\author[0000-0002-1832-8229]{Anne S.M. Buckner}
\affiliation{Cardiff Hub for Astrophysics Research and Technology (CHART), School of Physics \& Astronomy, Cardiff University, The Parade, CF24 3AA Cardiff, UK}
\email{BucknerA@cardiff.ac.uk}

\author[0000-0002-5189-8004]{Daniela Calzetti}
\affiliation{Department of Astronomy, University of Massachusetts Amherst, 710 North Pleasant Street, Amherst, MA 01003, USA}
\email{calzetti@astro.umass.edu}

\author[0000-0001-6464-3257]{Matteo Correnti}
\affiliation{INAF Osservatorio Astronomico di Roma, Via Frascati 33, 00078, Monteporzio Catone, Rome, Italy}
\affiliation{ASI-Space Science Data Center, Via del Politecnico, I-00133, Rome, Italy}
\email{matteo.correnti@inaf.it}

\author[0000-0002-5259-4774]{Ana Duarte-Cabral}
\affiliation{Cardiff Hub for Astrophysics Research and Technology (CHART), School of Physics \& Astronomy, Cardiff University, The Parade, CF24 3AA Cardiff, UK}
\email{adc@astro.cf.ac.uk}

\author[0000-0002-3247-5321]{Kathryn~Grasha}
\altaffiliation{ARC DECRA Fellow}
\affiliation{Research School of Astronomy and Astrophysics, Australian National University, Canberra, ACT 2611, Australia}  
\email{kathryn.grasha@anu.edu.au}

\author[0000-0001-8348-2671]{Kelsey Johnson}\affiliation{Department of Astronomy, University of Virginia, Charlottesville, VA, USA}
\email{kej7a@virginia.edu}

\author[0009-0009-5509-4706]{Drew Lapeer}
\affiliation{Department of Astronomy, University of Massachusetts, 710 North Pleasant Street, Amherst, MA 01003, USA}
\email{dlapeer@umass.edu}

\author[0000-0003-1427-2456]{Matteo Messa}
\affiliation{INAF–OAS, Osservatorio di Astrofisica e Scienza dello Spazio di Bologna, via Gobetti 93/3, I-40129 Bologna, Italy}
\email{matteo.messa@inaf.it}

\author[0000-0002-3005-1349]{G\"oran \"Ostlin}
\affiliation{Department of Astronomy, Oskar Klein Centre, Stockholm University, AlbaNova University Centre, SE-106 91 Stockholm, Sweden}
\email{ostlin@astro.su.se}

\author[0009-0002-2481-2217]{Linn Roos}
\affiliation{Max-Planck-Institut für Radioastronomie, Auf dem Hügel 69, 53121 Bonn, Germany}
\email{lroos@mpifr-bonn.mpg.de}

\author[0000-0003-2954-7643]{Elena Sabbi}
\affiliation{Gemini Observatory/NSFs NOIRLab, 950 N. Cherry Ave., Tucson, AZ 85719, USA}
\email{elena.sabbi@noirlab.edu}

%% Use the \collaboration command to identify collaborations. This command
%% takes an optional argument that is either a number or the word ``all"
%% which tells the compiler how many of the authors above the command to
%% show. For example ``\collaboration[all]{(DELVE Collaboration)}" wil include
%% all the authors above this command.
%%
%% Mark off the abstract in the ``abstract'' environment. 
\begin{abstract}

JWST can pierce through dusty molecular clouds to study the early stages of star formation, where young star clusters are actively driving stellar feedback and still emerging from their natal cloud. We present a first look of the JWST/NIRSpec multiplex spectroscopy observations acquired by the Feedback in Emerging extrAgalactic Star clusTers (FEAST) program for the nearby spiral galaxy NGC628. We showcase JWST's ability to resolve the spectral properties of emerging young star clusters (eYSCs) and their immediate interstellar medium (ISM) by focusing on a bright star-forming complex ($0.5\times0.5~\mathrm{kpc}^2$) in the northern spiral arm as a science proof-of-concept. The eYSC spectra are rich in ionized gas (from \hii regions), as well as warm H$_2$ and polycyclic aromatic hydrocarbon (PAH) emission from photodissociation regions (PDRs), consistent with young star formation. $\mathrm{Pa}\alpha$ equivalent widths and H/He ionizing photon fluxes both indicate the presence of hot, young massive stars (O8.5V-O8V), consistent with photometry SED estimates. The ionized gas is highly correlated with H$_2$ and PAH emission, suggesting that the PDR morphology evolves as clusters emerge from their natal cloud. We find a photoionization-dominated regime from independent line diagnostics, with little contribution from Supernovae-driven shocks, highlighting the importance of pre-Supernovae feedback when massive stars are present. This pilot study showcases how JWST's multiplex spectroscopy mode can disentangle the mechanisms present in the youngest stages of star formation for the first time outside the Local Group.

\end{abstract}

%% Keywords should appear after the \end{abstract} command. 
%% The AAS Journals now uses Unified Astronomy Thesaurus (UAT) concepts:
%% https://astrothesaurus.org
%% You will be asked to selected these concepts during the submission process
%% but this old ``keyword" functionality is maintained in case authors want
%% to include these concepts in their preprints.
%%
%% You can use the \uat command to link your UAT concepts back its source.

%% From the front matter, we move on to the body of the paper.
%% Sections are demarcated by \section and \subsection, respectively.
%% Observe the use of the LaTeX \label
%% command after the \subsection to give a symbolic KEY to the
%% subsection for cross-referencing in a \ref command.
%% You can use LaTeX's \ref and \label commands to keep track of
%% cross-references to sections, equations, tables, and figures.
%% That way, if you change the order of any elements, LaTeX will
%% automatically renumber them.

\section{Introduction}
\label{sec:introduction}

The transformation of interstellar gas into stars, and the subsequent stellar feedback, play a critical role in the baryon cycle of the Universe. The star formation process as a whole drives chemical abundances, dictates galaxy evolution, and sets the stage for planet formation \citep[e.g.,] []{schinnerer_2024ARA&A..62..369S}. Given its importance, it is crucial to understand how star formation unfolds, specifically how the different stellar feedback mechanisms affect the interstellar medium (ISM). 

Most stars form in clusters \citep{lada_2003ARA&A..41...57L} within cold and dense molecular clouds in the ISM \citep{bergin_2007ARA&A..45..339B}. Massive stars (\mbox{$\mathrm{M}>8~\mathrm{M}_\odot$}) within these young star clusters (YSCs) are the primary drivers of stellar feedback \citep{krumholz_2019_araa}. These processes inject energy, momentum and metals into the ISM, which can deter any further gas collapse and thus regulate the local star formation efficiency \citep[e.g.,] []{krumholz_2019_araa,grudic_2021}. Indeed, stellar feedback is often invoked as the main driver behind the long molecular gas depletion times observed at both cloud- and galactic-scales in nearby galaxies \citep[$1-2~\mathrm{Gyr}$;][]{bigiel_2011}, which are much longer than expected if the gas was freely collapsing under self-gravity to form stars \citep{zuckerman_evans_1974}.  

Recent results highlight the importance of pre-Supernovae feedback (pre-SNe; e.g., photoinization, stellar winds) in already dispersing the molecular clouds and regulating star formation \citep[for a review see][]{chevance_2023ASPC..534....1C}, which is further amplified with the advent of SNe explosions. This is supported both by observations \citep{mcleod_2019MNRAS.486.5263M,mcleod_2021MNRAS.508.5425M,barrera-ballesteros_2021MNRAS.503.3643B,pedrini_2024ApJ...971...32P,pathak_2025ApJ...982..140P,knutas_2025ApJ...993...13K} and simulations \citep{dale_2012MNRAS.427.2852D,skinner_2015ApJ...809..187S,jeffreson_2021MNRAS.505.3470J,grudic_2021,grudic_2022MNRAS.512..216G,ali_2023MNRAS.524..555A,andersson_2024A&A...681A..28A,neralwar_2025arXiv251007393N}. Still, the specific effects of the various stellar feedback mechanisms on the progenitor molecular clouds remain poorly understood \citep{schinnerer_2024ARA&A..62..369S}. 

To determine the impact of stellar feedback it is therefore crucial to target the early stages of star formation. In this early phase, the feedback-driving \textit{emerging} young star clusters (eYSCs) still coexist with the feedback-affected natal clouds. Massive stars ionize the surrounding gas, powering expanding \hii regions that emit bright H recombination lines \citep{draine_2011piim.book.....D} observable in the near-infrared (NIR) for eYSCs and optical-ultraviolet (UV) for exposed star clusters. Furthermore, photodissociation regions \citep[PDRs;][]{hollenbach_1999RvMP...71..173H} sit at the interface between the \hii regions and the molecular clouds. Within the PDRs, polycyclic aromatic hydrocarbons \citep[PAHs; for a review see][]{tielens_2008ARA&A..46..289T} reprocess UV photons into IR emission \citep[e.g.,] []{churchwell_2009PASP..121..213C,relano_2009ApJ...699.1125R,maragkoudakis_2020MNRAS.494..642M}. 
%PDRs are particularly visible with the PAH emission at $3.3~\mu$m arising from the C-H stretching vibration of the smallest and neutral PAHs, which are particularly susceptible to the radiation field \citep[e.g.,] []{maragkoudakis_2020MNRAS.494..642M,peeters_2024A&A...685A..74P,schroetter_2024A&A...685A..78S}. 
At the youngest evolutionary stages of eYSCs, all the components of the star formation cycle (i.e., \hii regions, PDRs) and associated stellar feedback signatures are obscured by dust.

The James Webb Space Telescope (JWST) is perfectly placed to pierce through the dusty molecular clouds and reveal the emerging clusters, which are still partly embedded and actively driving feedback. The wavelength coverage of JWST allows us to simultaneously capture the ionized gas properties (through NIR H recombination and He lines), the PDR properties (through PAH and warm molecular gas emission), and access emission line diagnostics that probe the nature of the stellar feedback mechanisms at play \citep[e.g.,] []{cresci_2010A&A...520A..82C,hunt_2025ApJ...993...84H}. With JWST's revolutionary resolution and sensitivity, it is now possible to conduct such studies of the early stages of star formation in galaxies beyond the Local Group. While studying star-forming regions in the Milky Way grants sufficient resolution to disentangle the individual components of YSCs \citep[e.g.,] []{zucker_2023ASPC..534...43Z} and associated feedback, we are limited by line-of-sight confusion to individual regions, and are not able to collect a full, unbiased census of YSCs. Local Group galaxies instead are limited by their low star formation rates, which result in a small number of massive clusters \citep{wainer_2022}. Nearby galaxies ($\mathrm{D}\leq10$~Mpc) provide the perfect opportunity to probe the entire star cluster population with minimal projection effects for face-on inclinations, and thus study the effects of stellar feedback on the ISM in a statistically meaningful way at parsec-scales \citep[e.g.,] [\textit{subm.}]{whitmore_2023,linden_2023,levy_2024,knutas_2025ApJ...993...13K,henny_2025ApJ...991...76H,pedrini_2025ApJ...992...96P}.

In this work, we use the multi-object spectroscopy (MOS) mode of JWST/NIRSpec \citep{nirspec_2022A&A...661A..80J} to unveil the properties of eYSCs and their immediate ISM in NGC\,628. These observations were acquired as part of the Feedback in Emerging extrAgalactic Star clusTers (FEAST; PI Adamo)\footnote{\href{https://feast-survey.github.io}{feast-survey.github.io}} survey. NGC\,628 is a spiral galaxy situated at 9.84~Mpc \citep{tully_2009AJ....138..323T}, with a face-on inclination \citep[$8.9^\circ$,][]{lang_2020ApJ...897..122L}, a global star formation rate of $1.7~\mathrm{M}_\odot\mathrm{yr}^{-1}$ \citep{leroy_2021}, and near-solar metallicity with a modest gradient \citep{berg_2020ApJ...893...96B}. 
The spectral capabilities of JWST allow us to disentangle the overlapping physical processes occurring at the youngest phases of star formation, rather than obtaining only their integrated effect \citep[i.e., photometric-based estimates; e.g.,] []{donnelly_2025ApJ...983...79D}. With JWST, we can now unlock this type of resolved, detailed analysis of the star formation cycle outside the Local Group.

This paper is organized as follows. Section~\ref{sec:data} describes the observations, the data reduction strategy, and how the analyzed properties are extracted from the spectra. Section~\ref{sec:results} outlines our findings on the spectral properties of eYSCs, PDRs, the diffuse ISM, and the active stellar feedback mechanisms. These results are further discussed in Section~\ref{sec:discussion}. We provide a summary of our conclusions in Section~\ref{sec:conclusion}.

 \begin{figure*}
    \centering
    \includegraphics[width=\textwidth]{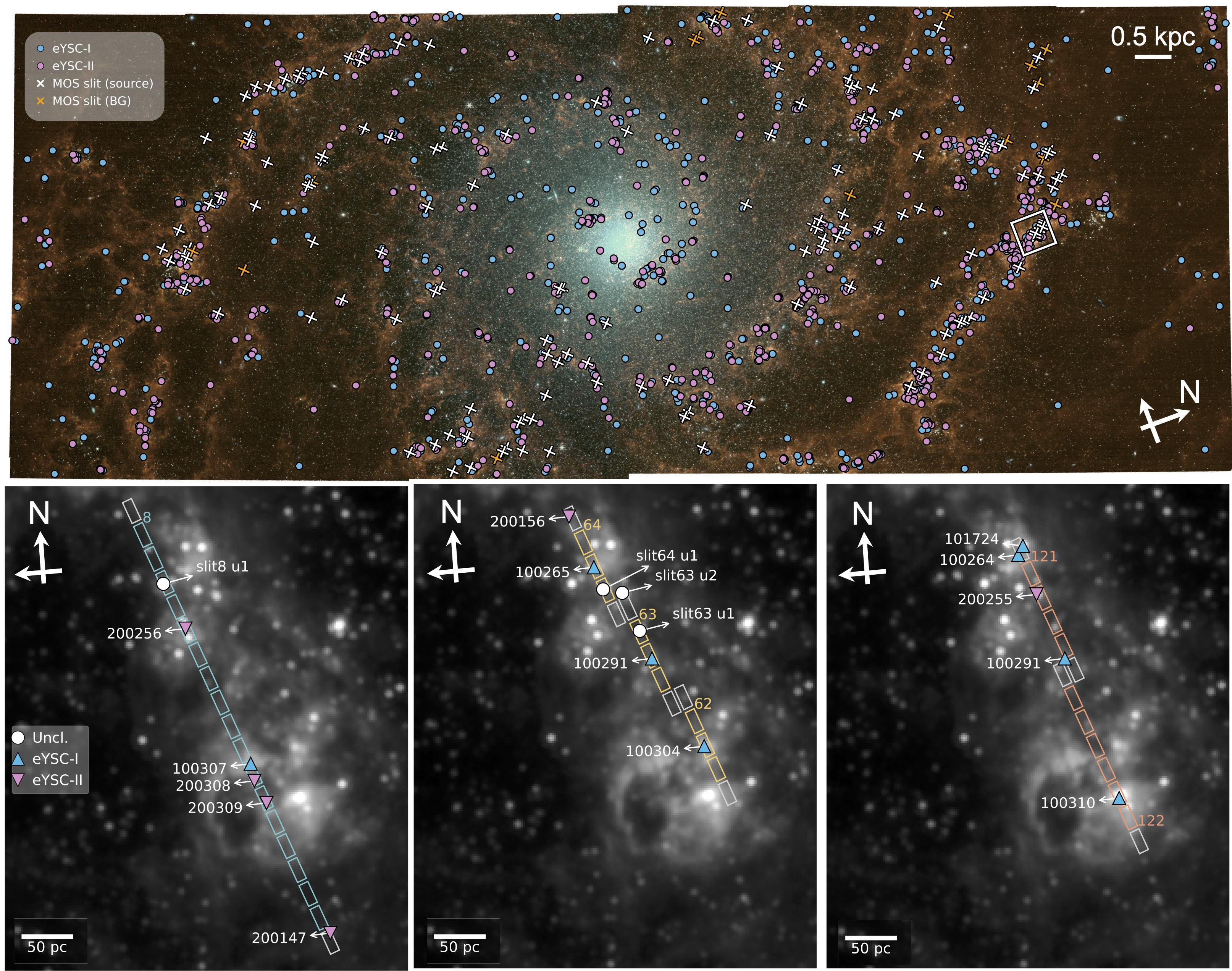}
    \caption{\textit{Top:} Composite image of NGC\,628 with the 3.3~$\mu$m PAH feature (NIRCam/F335M) in red, Br$\alpha$ (NIRCam/F405N) in green, and the stellar continuum (NIRCam/F444W) in blue. The circles denote the positions of the emerging young star clusters (blue for eYSC-I and pink for eYSC-II) from the FEAST catalog \citep[e.g.,] []{pedrini_2025ApJ...992...96P}. The crosses display the locations of all MOS slits targeting a source (white) or background (orange). The region of interest, which the present paper focuses on, is highlighted by the white box, and further zoomed into in the bottom 3 subplots. \textit{Bottom:} The zoomed-in subplots showcase the NIRCam/F335M in the background, and the slit footprints of the NIRSpec/MOS observations for this region (with slit IDs), across the 3 different MSA configurations (left, middle, right). The locations of eYSC-Is, eYSC-IIs and unclassified sources (see main text) within the spectroscopic slit observations are shown as blue triangles, pink triangles and white circles, respectively. The gray rectangles above and below each slit represent the additional shutters opened for the three-nod dithering pattern.}
    \label{fig:fullngc628}
\end{figure*}

\section{Observations and Data reduction}
\label{sec:data}

We present here a spectroscopic study of star-forming regions in the nearby spiral NGC\,628 (also known as M74). In Cycle 1, FEAST obtained NIRCam \citep{nircam_2023PASP..135b8001R} and MIRI \citep{miri_2023PASP..135d8003W} imaging across 10 bands (GO1783, PI Adamo), targeting the wavelength range $1.1-8~\mu$m. Figure~\ref{fig:fullngc628} displays a composite image of NGC\,628 made from 3 of the FEAST band observations: NIRCam/F444W tracing the underlying stellar and dust continuum, as well as NIRCam/F405N and NIRCam/F335M which are both continuum-subtracted to reveal the Br$\alpha$ and the 3.3~$\mu$m PAH emission respectively \citep[see][]{gregg_2024ApJ...971..115G} within the filters. More information on the survey and data processing can be found in the FEAST presentation paper (A. Adamo et al. \textit{in prep}), as well as other FEAST publications \citep[e.g.,][]{knutas_2025ApJ...993...13K,pedrini_2024ApJ...971...32P,pedrini_2025ApJ...992...96P,gregg_2024ApJ...971..115G,gregg_2025arXiv251106481G}.

In this work, we used observations of NGC\,628 taken with JWST/NIRSpec in the MOS \citep[][]{mos_2022A&A...661A..81F} mode, as part of the Cycle 2 FEAST program (GO3503, PI Adamo). The MOS mode uses the Micro-Shutter Assembly (MSA), where $\sim$250,000 microscopic shutters can be opened or closed to take simultaneous spectroscopic observations within a single exposure \citep{mos_2022A&A...661A..81F}. This is a powerful mode for nearby galaxy studies, as it allows us to obtain spectroscopic information for dozens of star-forming regions at high spatial resolution, at a relatively low cost. 

In this pilot study, we focus on one specific $\sim0.5\times0.5~\mathrm{kpc}^2$ region of NGC\,628, highlighted in Figure~\ref{fig:fullngc628}. This region is one of the brightest molecular cloud complexes in the galaxy, with 3.3~$\mu$m PAH emission permeating the entire region as well as a significant concentration of eYSCs, which is indicative of ongoing star formation. Furthermore, 6 different slits were opened towards this region, providing an ideal starting point to not only test our data reduction pipeline, but also to study signatures of stellar feedback in an active star-forming region. 
%Across the 6 slits in this region, we target 7 eYSC-Is and 6 eYSC-IIs. 
The goal of this work is, therefore, to present our data reduction strategy, initial results, and showcase the planned scientific analysis for the full spectroscopic dataset, which consists of a total of 165 slits. 

Our data products are available in the Mikulski Archive for Space
Telescopes (MAST) as a High Level Science Product via \url{https://doi.org/10.17909/pfmz-gx21} (DOI: 10.17909/pfmz-gx21) and on the FEAST webpage (\url{https://feast-survey.github.io/}).

\subsection{Source catalog}

In an effort to better understand the processes leading to the emergence of star clusters from their natal gas clouds, FEAST has identified young star clusters in its targeted galaxies, which include NGC\,628 \citep[A. Adamo et al. \textit{in prep};][]{pedrini_2025ApJ...992...96P}. This is done in both the NIR and the optical regime, in order to sample the cluster population from its embedded phases to the exposed, optically-bright stage.

The eYSCs are identified with JWST imaging, and are subdivided into 3 categories: (1) \textit{eYSC-I} show peaks of emission in both Pa$\alpha$/Br$\alpha$ and $3.3~\mu$m PAH, (2) \textit{eYSC-II} are detected with compact Pa$\alpha$/Br$\alpha$ emission but with more diffuse/extended PAH emission, and (3) \textit{$3.3~\mu$m PAH peaks} only show compact \pah emission but no compact H emission. For a more detailed description of the eYSC catalogs and their construction see \cite{knutas_2025ApJ...993...13K} and \cite{pedrini_2025ApJ...992...96P}. The optical star clusters are selected based on the HST/F555W filter. The methodology for constructing the optical YSC catalog for NGC\,628 is presented in \cite{knutas_2025ApJ...993...13K} and A. Adamo et al. (\textit{in prep}). We further required the optical YSCs to lie within the JWST field-of-view.

The combined NIR and optically-selected FEAST cluster catalogs paint the following evolutionary sequence: eYSC-Is (compact PAH) evolve into eYSC-IIs (no compact PAH), which in turn evolve into the fully exposed, optical young star clusters \citep{pedrini_2024ApJ...971...32P,knutas_2025ApJ...993...13K}. This is consistent with the expectation that the PDR-tracing 3.3~$\mu$m PAH feature \citep[which originates from the smallest and neutral PAH grains; e.g.,][]{maragkoudakis_2020MNRAS.494..642M} is quickly destroyed once ionization from massive stars kicks in \citep{guhathakurta_1989ApJ...345..230G,draine_2021ApJ...917....3D}. The exact position of the $3.3~\mu$m PAH peaks in this evolutionary sequence is yet to be determined, but their properties seem to be consistent with deeply embedded star clusters \citep[e.g.,][]{rodriguez_2023ApJ...944L..26R}, or with star clusters whose stellar population is not massive enough to power a detectable \hii region \citep{knutas_2025ApJ...993...13K}.

\subsection{MSA configuration}

The FEAST cluster catalogs were used as the foundation for the MSA planning and configuration of the NIRSpec/MOS observations outlined in this work. We applied luminosity cuts to the catalogs to ensure that the selected clusters would be detected with a signal-to-noise $\mathrm{S/N}\geq 5$ on the weakest emission line targeted (H$_2~1-0~S(1)$), in each single exposure. For both the eYSC and optical YSC catalogs, this corresponds to a magnitude cut of 23~ABmag to the continuum level measured in the F200W band. For the eYSCs, we applied an additional cut at 23~ABmag to the fluxes measured in the F187N (Pa$\alpha$) and F335M ($3.3~\mu$m PAH band) filters. The optical YSC catalog was further cut with an upper age limit of 20~Myr (determined from spectral energy distribution, SED, fitting; see \citealt{knutas_2025ApJ...993...13K,pedrini_2025ApJ...992...96P}) to maximize the sampling of luminous red supergiant (RSG) clusters. We applied weights to these cluster catalogs to optimize the amount of eYSCs (primary targets) observed for our observational setup, with optical YSCs constituting the filler targets. Within the primary targets, the weighting process prioritizes eYSC-Is, eYSC-IIs, and then \textit{$3.3~\mu$m PAH peaks}, as a function of their increasing luminosity in the F200W, F187N and F335M filters. The weighting on the filler sources was done to prioritize brighter and younger sources.

\begin{figure*}
    \centering
    \includegraphics[width=0.8\textwidth]{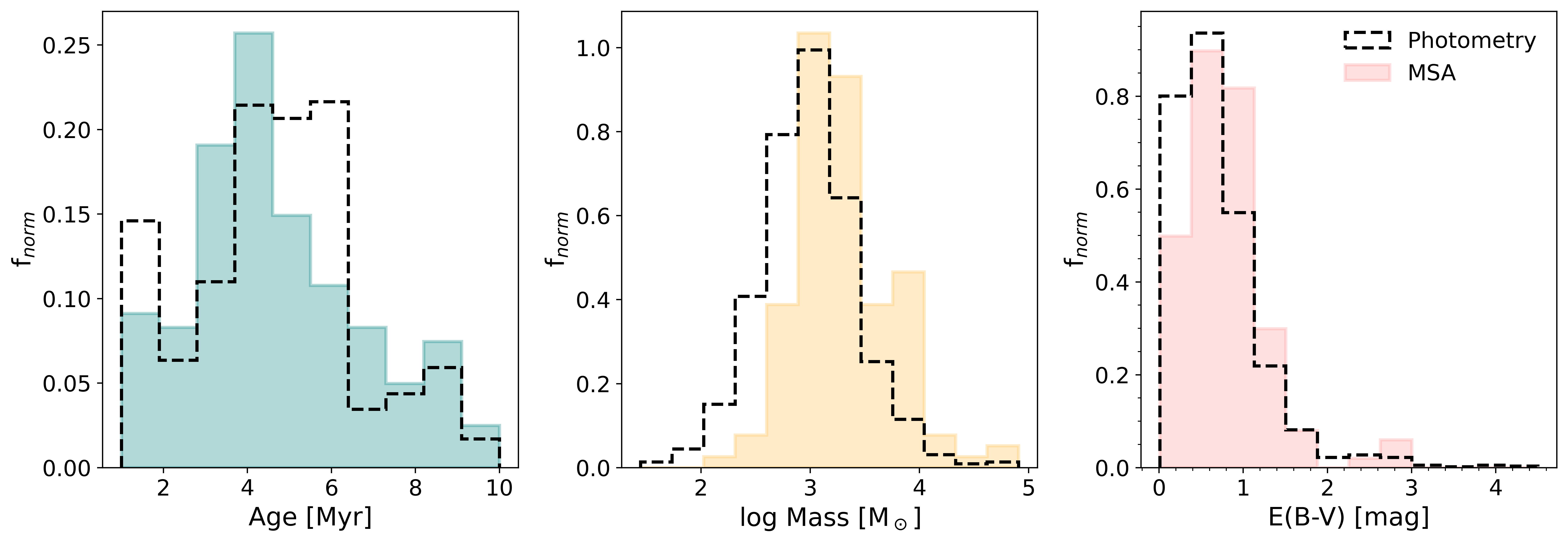}
    \caption{Normalized distributions of age (left), mass (middle), and extinction (right) for all FEAST eYSCs in NGC\,628 (dashed line) against the MSA targets (filled), from the \texttt{CIGALE} SED fits \citep[see text;][S. T. Linden et al. \textit{in prep}]{pedrini_2025ApJ...992...96P}. The MSA targets have properties which are representative of the full galaxy eYSC population.}
    \label{fig:msa_vs_phot}
\end{figure*}

The MSA planning tool was used to find the initial configuration of the mask with an assigned position angle (PA) of 25$^\circ$, which was found to maximize the number of targets observed. During the planning, we allowed for a 3-shutter nodding and midpoint pointing precision. Using a search grid of $20^{\prime\prime}$, we found that we were able to recover a consistent sampling of the eYSC population in NGC\,628 (see Figure~\ref{fig:msa_vs_phot}) with 3 MSA configurations. The slight shift towards higher masses in the MSA subsample relative to the full photometric eYSC sample (shown in Figure~\ref{fig:msa_vs_phot}) is a natural consequence of the luminosity cuts applied when optimizing the catalog to ensure high S/N observations. The initial sample consists of 155 primary and 36 secondary targets. In a subsequent intervention, the MSA masks were manually improved/optimized to allow a better coverage of specific regions. To maximize the target sample, we also allow slits with only 2 shutters (in this case, the targets would be observed only with 2 exposures) in the few cases where the standard 3-shutter slit would cause the resulting spectrum to overlap with another source or coincide with closed shutters. Additionally, we opened ``background-only" slits in each MSA configuration to sample the average galactic background (see Sec.~\ref{sec:sky_bg}). As can be seen from the bottom panels of Figure~\ref{fig:fullngc628}, our slit configurations vary in the number of open shutters in an effort to encompass as much of the targeted star-forming region as possible. Longer slits enable the study of the variation of ISM tracers across these regions, as a function of distance from ionizing sources (see Sec.~\ref{sec:ism_spectra}).

%We built 3 MSA masks (i.e., open/closed shutter configurations) that span the full field-of-view of the NIRCam imaging based on the photometric eYSC catalog from FEAST described above. 
Across the 3 MSA configurations, we placed 165 open slits. Slits are composed of contiguous open shutters (see bottom row of Figure~\ref{fig:fullngc628}) of size $0.2^{\prime\prime}\times0.46^{\prime\prime}$ (or $9.5\times21.9$~pc), with a small spacing of $0.07^{\prime\prime}$ (3~pc) between shutters. Of these slits, 146 target eYSCs, while 19 are ``background-only" slits, which are important for our sky subtraction strategy (Sec.~\ref{sec:sky_bg}). In total, 96 eYSC-Is, 43 eYSC-IIs, 16 3.3~$\mu$m peaks (155 eYSCs in total) are targeted, as well as 36 optical YSCs. In many cases, multiple sources will be observed by the same slit. In this work, source IDs in the format of $1\mathrm{XXXXX}$ belong to the eYSC-I class, and $2\mathrm{XXXXX}$ to eYSC-II. We also denote ``unclassified" sources (i.e., sources not within the photometric catalog, see Sec.~\ref{sec:source_spectra_extr}) as ``slit\# $u$X", where \# refers to the slit ID.

We observed all 3 of our MSA configurations with the following medium-resolution disperser/filter combinations: G140M/F100LP, G235M/F170LP, and G395M/F290LP. This spectroscopic setup allows for a wavelength coverage from 0.97 to 5.29~$\mu$m at $\mathrm{R}\sim1000$, ensuring that we detect key emission lines for the study of young star-forming regions, such as He~I, several of the H recombination lines, [Fe II], H$_2$, and the PAH feature at 3.3~$\mu$m. We used the NRSIRS2RAPID readout pattern, with 31, 32 and 34 groups per integration for the G140M, G235M and G395M gratings, respectively. This resulted in integration times of $1400$~s (G140M), $1444$~s (G235M), and $1532$~s (G395M), for each single nodding position in each MSA configuration. We outline our data reduction process in the following subsections.

\subsection{Data reduction}
\label{sec:reduction_pipeline}

The NIRSpec/MOS data presented here were processed using the 1.17.1 version of the JWST calibration pipeline \citep{jwstpipeline_2022zndo...7229890B}, with CRDS (Calibration Reference Data System) version 12.0.9 and context jwst\_1322.pmap. The pipeline is composed of three stages: instrumental artifact corrections (Stage 1), spectroscopic calibration (Stage 2), and spectral combination (Stage 3). We turned off any background subtraction performed by the JWST pipeline, opting instead to perform our own custom ``sky" removal (see Sec.~\ref{sec:sky_bg}). Additionally, given the significant amount of extended emission within the slit observations, we also performed our own 1D spectrum extraction from the calibrated and rectified Stage 3 outputs with custom apertures (Sec.~\ref{sec:spectra_extr}).

Stage 1 removes detector-level artifacts from the uncalibrated raw exposures. Bad pixels (e.g., hot or saturated pixels) were flagged. Detector drifts were accounted for through a reference pixel correction. The master bias frame and dark current were subtracted from the image. The count-rates of each exposure were ramp-fitted in order to detect any jumps caused by cosmic ray events, which were then flagged and rejected at a $4\sigma$ level. Additionally, to better handle large cosmic ray events (``snowballs"), we allowed the pipeline to expand the jump flagging to the neighboring pixels around the snowballs if they encompass a minimum of 5 contiguous ``jump-flagged" pixels. We also corrected the correlated $1/f$ noise introduced by the detector readout system due to thermal instabilities. This manifests as vertical banding in the 2D count-rate images, and if left uncorrected leads to wavelength-dependent undulations in the final spectra. We used the \texttt{clean\_flicker\_noise} (CFN) step in Stage 1 to address this noise. The background in each count-rate image is modeled by CFN using iterative sigma-clipping, and any pixels above a $2\sigma$ threshold from the median value in the image are considered ``signal" and disregarded from the background modeling. Bad pixels were also not included in the background modeling. The residual noise in the background is fitted with a median along the detector columns (axis=0), and subsequently removed from the count-rate image in an iterative fashion. After visual inspection of all $1/f$-corrected count-rate images and resulting residuals, we found that CFN sufficiently mitigates the vertical banding without over-subtracting the background and significantly altering the flux within the open slit footprints, which we found occurred occasionally with \texttt{NSClean}\footnote{\url{https://jwst-pipeline.readthedocs.io/en/stable/jwst/nsclean/main.html}}, the $1/f$ correction implementation in Stage 2 of the pipeline.

With Stage 1 complete, the count-rate images were put through Stage 2 of the pipeline. Positions in the detector frame were converted to physical world coordinates (both spatial and wavelength). The exposures were then flat-fielded to account for variations in the pixel-to-pixel sensitivity. The signal losses along the optical path were corrected as a function of wavelength using calibration reference files from CRDS. This step also attempts to account for slit losses due to the position of the source within the MSA shutter, but it was found to be insufficient. We corrected for this flux loss at a later stage (Sec.~\ref{sec:source_spectra_extr}, see Appendix~\ref{app:msafit}). Particularly important for extended sources such as ours, the data was corrected for the MSA ``bar shadow" (caused by the $0.07^{\prime\prime}$ spacing between shutters) using CRDS calibration reference files. For each wavelength point, we converted from counts/s to $\mathrm{MJy}~\mathrm{sr}^{-1}$. As mentioned above, we performed no background subtraction at this stage. 

\begin{figure*}
    \centering
    \includegraphics[width=0.8\textwidth]{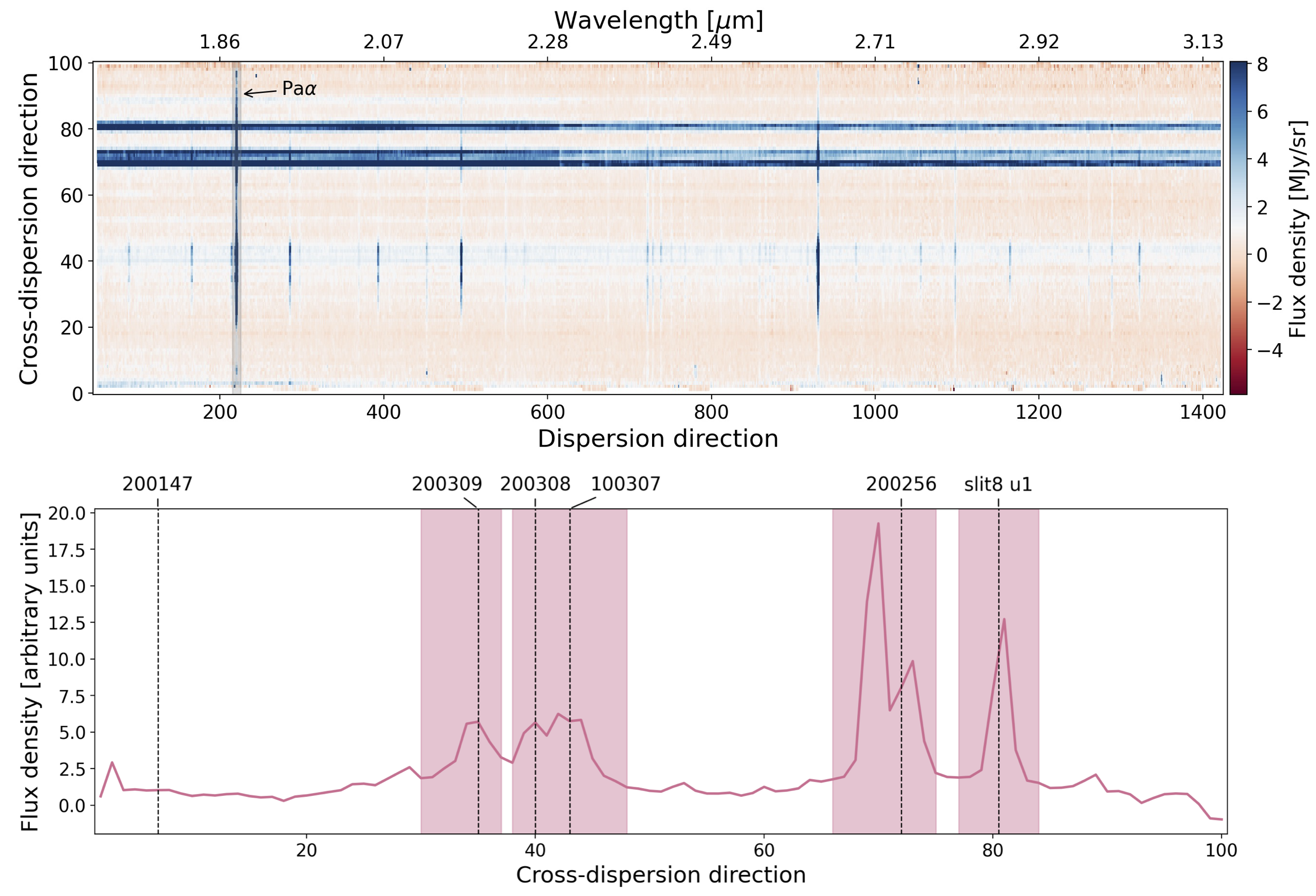}
    \caption{Example of the methodology adopted for the extraction of source spectra. \textit{Top:} 2D spectrum of the G235M grating for slit 8 (see Figure~\ref{fig:fullngc628}). The corresponding wavelengths (in $\mu$m) to the dispersion pixels are shown in the top x-axis. The Pa$\alpha$ line at $1.87~\mu$m is highlighted. \textit{Bottom:} Median cross-dispersion profile around the Pa$\alpha$ line for slit 8 (taken within the gray region in the top subplot, see main text). The dashed lines display the position of the cataloged eYSCs and ``unclassified" sources in this slit. The pink shaded regions denote the extraction aperture (i.e., range of cross-dispersion pixels) chosen for the relevant sources.}
    \label{fig:s2d}
\end{figure*}

Finally, in Stage 3, the calibrated data from multiple exposures (i.e., different dithers) were aligned, rectified, resampled onto a common grid, and combined into a single 2D image per slit for each grating (i.e., 2D spectrum, see Figure~\ref{fig:s2d}). The 2D spectra have spectral information in the dispersion direction (x-axis) and spatial information in the cross-dispersion direction (y-axis). 

\subsection{Spectrum extraction}
\label{sec:spectra_extr}

The 1D spectra are directly extracted from the combined, rectified 2D spectra (i.e., output of Stage 3 of the pipeline) using custom apertures. These encompass the full dispersion axis (i.e., x-axis of 2D spectra, see Figure~\ref{fig:s2d}), and a selected range of cross-dispersion pixels (y-axis). We outline our spectral extraction strategy in the following subsections.

\subsubsection{Background ``sky" spectra}
\label{sec:sky_bg}

The most common method of background subtraction used for NIRSpec/MSA data consists of calculating the ``local" sky background from the ``background" shutters within a slit \citep[e.g.,] []{bunker_2024A&A...690A.288B}. In other words, in a standard 3-shutter slit configuration, the source is targeted by one shutter, and the 2 remaining shutters observe background. Since we are not only interested in the spectral properties of the eYSC, but also the extended \pah emission permeating the star-forming region, performing a local background subtraction does not suit our objectives, given that it will likely over-subtract the extended \pah emission. Additionally, we often have different sources in multiple shutters in a single slit, reducing the amount of available ``background" shutters. In this work, we instead created ``global" background spectra for each MSA configuration using the 19 background-only slits in our dataset. These slits were opened in regions of NGC\,628 that are devoid of bright emission, specifically for this purpose. These background-only slits are not, of course, picking up a pure sky signal, but are also detecting the large-scale, diffuse emission from the galaxy. We visually inspected the 2D spectra of each of the 19 background-only slits, and selected apertures which do not include any obvious emission lines or strong continuum (which are indicative of sources), or strong \pah emission. In this process, 2 slits were disregarded due to significant source contamination. We extracted 1D spectra (each grating is handled separately) for each background-only slit within the chosen apertures. Across the different slits, the wavelength range of the observations vary slightly, since the slits themselves fall in different locations in the detector. Thus, we resampled all the background 1D spectra onto a common wavelength frame for each grating\footnote{Similar to what is done in the master background step in Stage 2 of the pipeline.}.

Since the 3 different MSA configurations were observed at different times, we calculated the background spectra for each configuration separately to ensure a robust measure of the sky conditions at the time. So, for each MSA configuration, we constructed the global background from the median flux density of the background-only slit spectra, at each wavelength point. We estimated the uncertainty of the background spectra from the median absolute deviation (MAD), as this is a measure of the scatter around the median value that is robust to outliers. The MAD is defined as \mbox{$\mathrm{MAD}=\left(|f(\lambda)_i - f(\lambda)_\mathrm{med}|\right)_\mathrm{med}$}, where $f(\lambda)_i$ is the flux density at wavelength $\lambda$ for slit $i$, and $f(\lambda)_\mathrm{med}$ is the median flux density value for $\lambda$ considering all background slits. At each wavelength, the MAD is consequently the median value of the absolute difference between the flux densities of each background slit to the median flux density. The intrinsic uncertainties of the flux densities of each background-only slit are already captured in the MAD, and thus we do not further include them in our error calculations. Finally, the background spectra were converted from $\mathrm{MJy}~\mathrm{sr}^{-1}$ to \erg.

\subsubsection{Source spectra}
\label{sec:source_spectra_extr}

We extract our source spectra from the Stage 3 pipeline outputs. In this process, the choice of aperture, i.e., the amount of cross-dispersion pixels considered, is based on the peaks of the Pa$\alpha$ line in the cross-dispersion direction. This is a reasonable approach, given that the FEAST eYSCs are identified from bright and compact Pa$\alpha$ emission. The Pa$\alpha$ cross-dispersion profile, shown in the bottom subplot of Figure~\ref{fig:s2d}, was constructed by taking all the spectral pixels within an observed wavelength range of $0.01~\mu\mathrm{m}$ from the Pa$\alpha$ central wavelength (1.87~$\mu$m) in the G235M 2D spectrum\footnote{We avoid using the few first and final pixels of gratings given the increase in uncertainty, and thus the Pa$\alpha$ line is taken from G235M rather than G140M.}, and retrieving the median along the cross-dispersion axis (shaded region in top panel of Figure~\ref{fig:s2d}). The resulting Pa$\alpha$ profile of each slit was then visually inspected, and apertures were chosen to capture, where possible, the full spread of the Pa$\alpha$ peaks above the continuum level, without any overlap with another source. We used the coordinates of the eYSC photometry catalog for guidance in this step. In a final check, we also construct \pah cross-dispersion profiles to judge whether the Pa$\alpha$-selected aperture is including most of the PAH emission (which can be more extended than Pa$\alpha$, particularly for eYSC-IIs) associated with the star cluster. If not, the aperture is then adjusted accordingly where possible, avoiding overlaps with other sources in the slit. Any significant Pa$\alpha$ peaks which were not attributed to a cataloged source were also extracted if they displayed emission lines and/or continuum. In this paper, we refer to these sources as ``unclassified", and their IDs are written in format ``slit\# $uX$", where \# refers to the relevant slit ID. Since these unclassified sources are identified from the detector frame of the 2D spectrum images, their physical coordinates are an approximate location based on the coordinate transformation information stored in the pipeline data products. Given that their precise location is not known, we refrain from repeating the FEAST photometry-based eYSC categorization, and leave them as unclassified. Figure~\ref{fig:s2d} shows an example of an unclassified source (slit\,8~$u1$). Aperture sizes are preferentially set to be $7-9$~pix ($\sim0.73-0.94^{\prime\prime}$\footnote{The average size of a pixel in the cross-dispersion direction is $0.105^{\prime\prime}$ \citep{nirspec_2022A&A...661A..80J}.} or $34.9-44.9$~pc) wide, with an imposed minimum of 5~pix ($0.54^{\prime\prime}$ or 24.9~pc; see Appendix~\ref{app:msafit}).

In some cases, more than two eYSCs are observed in a single MSA shutter, and their signal becomes blended in the 2D spectra (e.g., 100307 and 200308 in slit 8, see Figure~\ref{fig:s2d}). For these instances, if the peaks in the Pa$\alpha$ cross-dispersion profile are too close together such that deblending is not feasible, we selected one single aperture and extracted the blended spectrum. In other cases, no significant Pa$\alpha$ signal is detected towards the location of an eYSC (e.g., 200147 in slit 8, see Figure~\ref{fig:s2d}). This is likely due to the slit configuration, where a source might fall on the edge of the slit and therefore is not observed by all 3 dithers.

The spectrum of each source in a slit was extracted by summing up all the flux density within the aperture for each wavelength point. This was again done for each grating separately. We estimated the uncertainties on the flux densities from the square root of the sum of the three variance arrays (Poisson, read noise and flat field, also outputs from the pipeline) within the extraction aperture. This is the same procedure as that adopted by the JWST calibration pipeline. 

%All our custom-extracted spectra (and uncertainties) for the cataloged eYSC sources match the pipeline outputs from the Stage 3 \texttt{extract\_1d} step.

\begin{figure*}
    \centering
    \includegraphics[width=\textwidth]{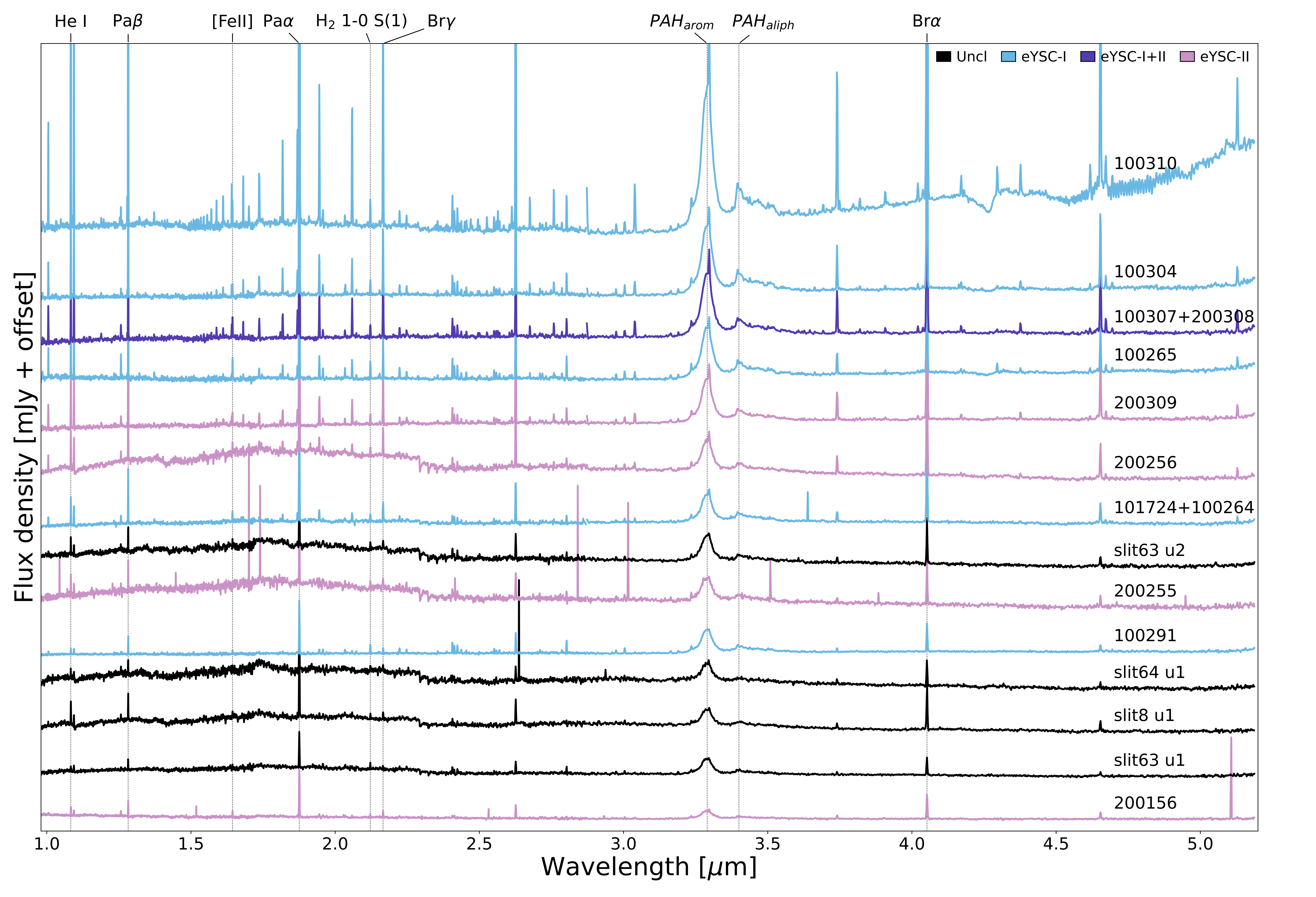}
    \caption{Spectra of sources within the region presented in this work (see Figure~\ref{fig:fullngc628}), ordered by decreasing \pah emission strength from top to bottom. The presented fluxes have been corrected for extinction, and wavelengths are in the rest-frame. The spectra are color-coded as: blue for eYSC-I, pink for eYSC-II, purple for blended eYSC-I+II, and black for unclassified sources. The spectra are offset by an arbitrary amount to avoid overlaps. The key emission lines and features used in the present analysis are highlighted by the dashed lines, and labeled.}
    \label{fig:source_spectra_all}
    
    \includegraphics[width=0.9\linewidth]{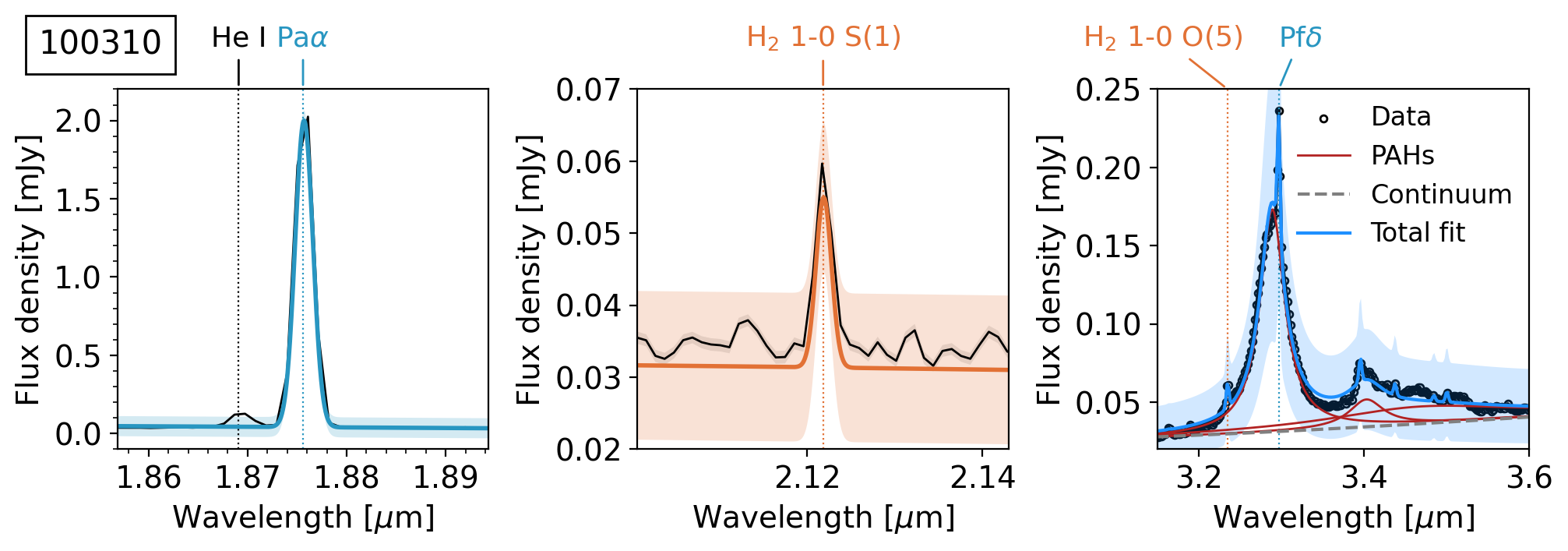}
    \caption{Examples of our spectral line fitting for source 100310. Single emission lines are fitted with a Gaussian profile, and \texttt{CAFE} is used for the \pah feature (see Section~\ref{sec:line_fluxes}). We show Pa$\alpha$ (\textit{left}), H$_2~1-0~S(1)$ (\textit{middle}) and the \pah feature (\textit{right}) as representative examples. The source spectrum is shown in black. The fit is shown as the solid colored line, with the shaded region corresponding to the standard deviation of the Monte Carlo iterations of the fits. For the right panel, we show \texttt{CAFE}'s continuum fit in the dashed gray line, and the PAH component fits in red (i.e., PAH$_{arom}$, PAH$_{aliph}$ and the plateau feature at $3.47~\mu$m). Emission line labels can be seen above each panel.}
    \label{fig:line_fits}
\end{figure*}

The background signal was subtracted from these source spectra, using the relevant configuration background spectrum (see Sec.~\ref{sec:sky_bg} above). The uncertainties from both the source and background flux densities are propagated in this subtraction. We applied an aperture correction on the extracted spectra based on forward-modeling performed with \texttt{MSAFIT} \citep{degraaff_msafit_2024A&A...684A..87D}, which also accounts for flux losses due to the source's position in the MSA shutter (see Appendix~\ref{app:msafit}). Additionally, we compare photometry measurements of the cataloged eYSCs to our spectra convolved with the relevant JWST/NIRCam filter throughput, and find a reasonable agreement between the spectroscopy and photometry estimates (see Appendix~\ref{app:msafit}). The extracted source spectra analyzed here are shown in Figure~\ref{fig:source_spectra_all}, ordered by \pah emission strength (with strongest at the top and weakest at the bottom). In this paper, we do not stitch the gratings together to form a single spectrum, and instead opt to use the single segments to do the proof-of-concept science. Regardless, as can be seen from Figure~\ref{fig:source_spectra_all}, we do not observe any evident flux disparities between the individual gratings across all eYSCs.

\subsubsection{ISM spectra}
\label{sec:ism_spectra_extr}

As is already evident from Figure~\ref{fig:fullngc628}, JWST allows us to observe and study not only the eYSCs still embedded in their natal cloud, but also their immediate environment that is full of diffuse \pah emission. Our MSA configuration masks were built with this in mind - to allow us to investigate the properties of the immediate ISM as a function of distance to the highly energetic and ionizing eYSCs. For that purpose, longer slits (i.e., more shutters were opened than the standard $1\times3$ design) were placed in particular star-forming hotspots, such as slit 8 shown in Figure~\ref{fig:fullngc628}. 

With this goal in mind, we again extracted ``ISM spectra" from the 2D spectra outputs from Stage 3 of the JWST pipeline. We selected equal apertures of 3~pix ($0.32^{\prime\prime}$ or 15~pc) in the cross-dispersion direction across the full 2D spectra, disregarding the first 5~pixels on either end of the slit, as these are typically associated with more noise. We did not exclude spectral pixels associated with eYSCs, since we are interested in the relative flux variation along the slit (and therefore across the star-forming region), and doing so would greatly reduce the amount of available apertures. In this work, we only performed this exercise for slit 8, as this is the longest slit in the region studied in this paper. Similar to the eYSC spectral extraction (see Sec.~\ref{sec:source_spectra_extr}), we summed up all the flux density per wavelength point within each aperture. The uncertainties were also estimated from the square root of the sum of the three noise variance arrays within each extraction aperture. The spectra were then background-subtracted, using the relevant MSA configuration background (see Sec.~\ref{sec:sky_bg}). Unlike the source spectra (see above), we did not perform any aperture correction here, since we are only interested in the relative differences of these ISM spectra as a function of distance to the ionizing eYSCs. Thus, the absolute values presented here for the ISM analysis should be handled with the appropriate care. This extraction results in 29 distinct ISM spectra for slit 8 (see Sec.~\ref{sec:ism_spectra}).

\subsection{Measured spectral features and uncertainties}
\label{sec:spectral_properties}

In this work, we study the spectral properties of emerging star clusters as well as their natal ISM. With critical tracers such as NIR H recombination lines (e.g., Paschen and Brackett series), He~I, [Fe\,II], warm H$_2$ and the \pah emission, we can probe the ongoing stellar feedback processes affecting the ISM as clusters begin to clear their molecular cloud. All steps outlined in this Section are applied to both the eYSC (Sec.~\ref{sec:source_spectra_extr}) and the ISM (Sec.~\ref{sec:ism_spectra_extr}) spectra.

\subsubsection{Continuum and line fitting}
\label{sec:line_fluxes}

In this paper, we focus on a selection of key stellar feedback and PDR tracers, which are highlighted in Figure~\ref{fig:source_spectra_all}. These include Pa$\alpha$ (at $1.87~\mu$m) and He~I (at $1.08~\mu$m) tracing photoionization, [Fe\,II] (at $1.64~\mu$m) which traces shocks, and the H$_2$~$1-0~S(1)$ line (at $2.12~\mu$m), which is one of the brightest warm molecular gas lines that can arise from PDRs. Finally, we also study the NIR PAH emission, paying particular focus to its different components: the aromatic component at $3.3~\mu$m (PAH$_{arom}$), and the aliphatic component at $3.4~\mu$m (PAH$_{aliph}$).

For each emission line, we first found adjacent spectral windows with no significant emission, and fitted the continuum with a first order polynomial (model from \texttt{astropy}). Considering the continuum fit, we then fitted the emission lines with a single Gaussian profile (from \texttt{astropy}) for all the key emission lines outlined above, except the PAH feature and the [Fe\,II] line. The fitting was performed with \texttt{astropy}'s damped least squares fitter function (\texttt{LevMarLSQFitter}). [Fe\,II] overlaps slightly with the Brackett transition H~$(12-4)$, or Br-12, (at $1.641~\mu$m), and thus we fitted this doublet with a tied, double Gaussian model with a fixed wavelength offset. We derived measurements for the PAH feature with \texttt{CAFE}\footnote{\url{https://github.com/GOALS-survey/CAFE}} \citep{cafe2025ascl.soft01001D}.

The uncertainties of the integrated line fluxes were estimated with a Monte Carlo approach. For each line flux measurement, we perturbed the spectrum using the flux density uncertainty (Sec.~\ref{sec:source_spectra_extr}). For each of the 100 perturbations, the relevant emission line is refitted following the methodology outlined above. The standard deviation of the resulting Monte Carlo measurements gives the uncertainty on the measured line fluxes. Examples of line fits (and uncertainties) are shown in Figure~\ref{fig:line_fits}. %

Finally, we corrected each spectrum for redshift (and velocity shifts due to galaxy rotation), using the average $z$ estimated from several $\lambda_\mathrm{obs}-\lambda_\mathrm{rest}$ measurements of the brightest H recombination lines (Pa$\alpha$, Pa$\beta$, Br$\alpha$ and Br$\beta$), where $\lambda_\mathrm{obs}$ is the observed peak wavelength (determined from the Gaussian fit) and $\lambda_\mathrm{rest}$ the rest-frame wavelength. We retrieve a median $z\sim0.00226$ for all the slits in this region of NGC\,628.

\subsubsection{Extinction correction}
\label{sec:ext_corr}

We estimated the extinction affecting each spectrum using the Br$\alpha$($4.05~\mu\mathrm{m}$)/Pa$\beta$($1.28~\mu\mathrm{m}$) line ratio. We chose these specific H lines given that they are sufficiently bright (we detect them in all our spectra), and encompass a large wavelength range. We calculated the intrinsic, theoretical emissivities of both lines assuming case B recombination, for an electron temperature $T_e=10^4$~K and an electron density $n_e=10^3~\mathrm{cm}^{-3}$ which are typical of \hii regions \citep{hummer_1987MNRAS.224..801H}. The color excess, $E(B-V)$, was measured by comparing the intrinsic line ratio, $R_\mathrm{int}$, to the observed ratio, $R_\mathrm{obs}$, following \cite{calzetti_2001PASP..113.1449C}:

\begin{equation}
    E(B-V) = \frac{2.5}{k\left(\mathrm{Br}\alpha\right)-k\left(\mathrm{Pa}\beta\right)} ~ \mathrm{log}\left(\frac{R_\mathrm{int}}{R_\mathrm{obs}}\right),
    \label{eqn:ebv}
\end{equation}

\noindent which assumes a foreground screen geometry. We determined $R_\mathrm{obs}$ by measuring the ratio of the Br$\alpha$ and Pa$\beta$ fluxes. For our chosen H line ratio, $R_\mathrm{int}\simeq0.487$ \citep{pyneb_2015A&A...573A..42L}. We used the $R_V=3.1$ model from \cite{gordon2023ApJ...950...86G}, which is optimized for JWST wavelengths, to compute the extinction curve values for Br$\alpha$ and Pa$\beta$, $k\left(\mathrm{Br}\alpha\right)=0.0358$ and $k\left(\mathrm{Pa}\beta\right)=0.2535$.

We estimated the uncertainty of the measured \mbox{$E(B-V)$} by first measuring the error on $R_\mathrm{obs}$. This was done by propagating the Monte Carlo errors of the Br$\alpha$ and Pa$\beta$ fluxes in their division (Eqn.~\ref{eqn:ebv}). This $R_\mathrm{obs}$ uncertainty was then used to estimate the error on \mbox{$E(B-V)$}, using a Monte Carlo approach (see Sec.~\ref{sec:line_fluxes}). %The extinction, $A_V$, was measured through $A_V = R_V \times E(B-V)$, with $R_V=3.1$.        

The spectra, both for sources and ISM, were corrected for extinction following \cite{calzetti_2001PASP..113.1449C}:

\begin{equation}
    F(\lambda) = F(\lambda)_\mathrm{obs} \times10^{0.4 ~ k(\lambda) ~ E(B-V)},
    \label{eqn:ext_corr}
\end{equation}

\noindent where $F(\lambda)$ is the extinction-corrected flux density for wavelength $\lambda$, $F(\lambda)_\mathrm{obs}$ the observed flux density, and $k(\lambda)$ the extinction curve value again measured with the \cite{gordon2023ApJ...950...86G} model for $R_V=3.1$. We were not able to reliably retrieve extinction estimates for 2 of our fainter sources (200156 and 200255), as the observed line ratio is very close to the $R_\mathrm{int}$ value. For these cases, we set $E(B-V)=0$ and applied no extinction correction. We were also not able to robustly measure extinction for some of the ISM spectra extracted from slit 8. In these cases, we applied no extinction correction and treat the resulting fluxes as lower limits (see Sec.~\ref{sec:ism_spectra}). For the sources studied in this work, we measure \mbox{$E(B-V)$} between $0.12-2.89$~mag.

\begin{figure*}
    \centering
    \includegraphics[width=\linewidth]{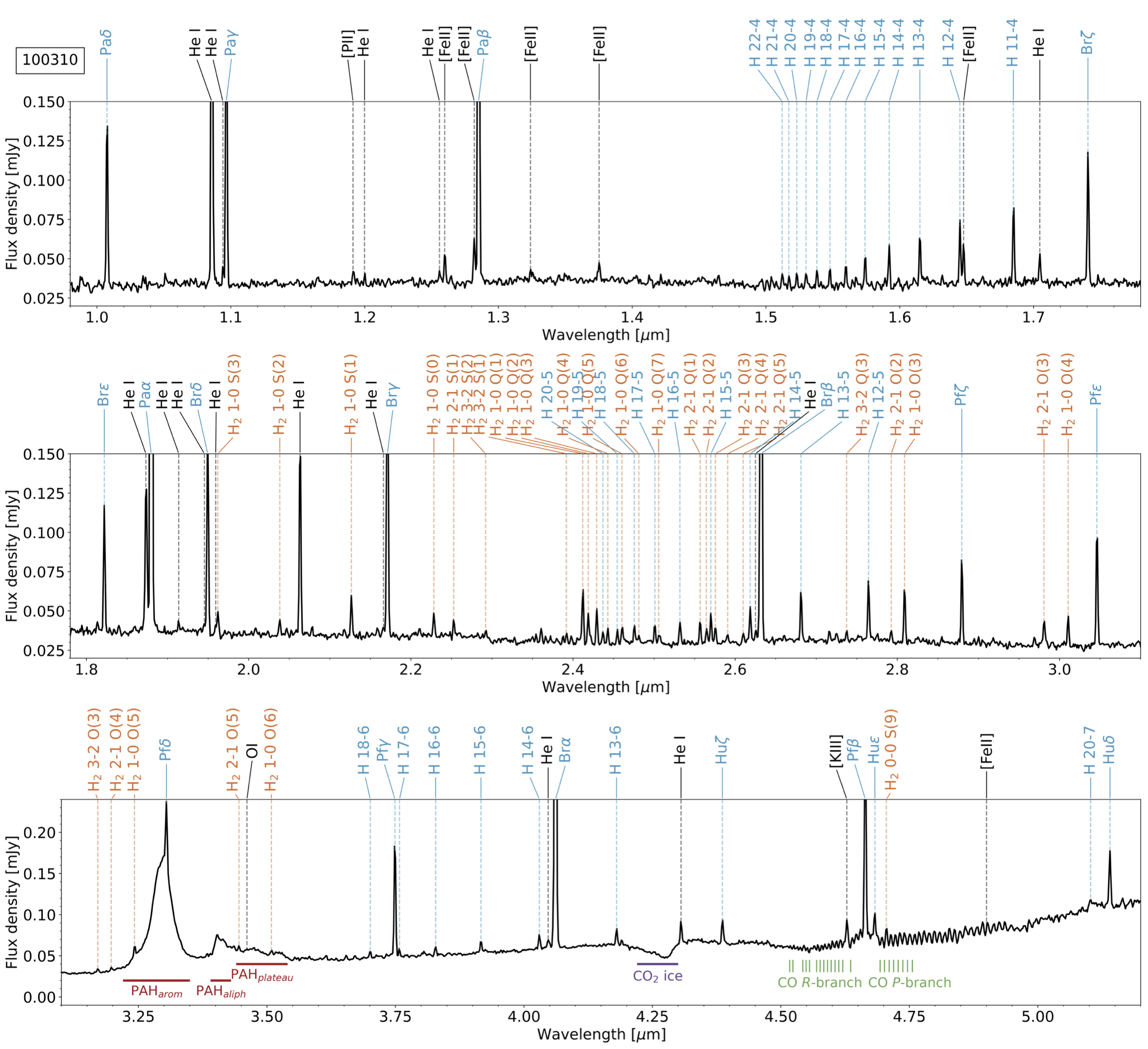}
    \caption{Spectrum of source 100310 (top spectrum in Figure~\ref{fig:source_spectra_all}) subdivided into 3 subplots to better illustrate the richness of emission lines and absorption features detected in the FEAST MOS observations. The spectrum shown in black is not continuum-subtracted, but is corrected for extinction and redshift. Identified lines are shown by the black dashed lines, with H recombination lines highlighted in blue, H$_2$ lines in orange, the PAH components in red, the CO$_2$ ice absorption feature (at 4.27~$\mu$m) in dark purple, and gas-phase CO transitions in the $R$ and $P$ branches in green (see text).}
    \label{fig:spectra_zoom}
\end{figure*}

\section{Results}
\label{sec:results}

Figure~\ref{fig:source_spectra_all} showcases the 14 initial eYSC spectra extracted from the FEAST NIRSpec/MOS observations of NGC\,628. The spectra show a richness of emission and absorption features at a level of detection that is unprecedented for a galaxy at this distance (see also Figure~\ref{fig:spectra_zoom}), showcasing JWST's unrivaled sensitivity and resolution. In fact, we reach high S/N for even the faintest sources detected: we measure S/N$\sim36$ for Pa$\alpha$, S/N$\sim14$ for Pa$\beta$ (used for extinction measurements, see Sec.~\ref{sec:ext_corr}), S/N$\sim4.5$ for Br$\gamma$, and S/N$\sim3$ for H$_2 \, 1-0 \, S(1)$. The quoted S/N values mentioned are calculated by dividing the amplitude of the Gaussian line fit by the standard deviation of the continuum flux measured in a similar sized spectral window close to the line.

Despite the relatively small size of the star-forming complex studied in this work ($\sim0.5\times0.5~\mathrm{kpc}^2$), the spectra show a wide variety of features. Figure~\ref{fig:spectra_zoom} shows the spectrum of the brightest eYSC studied here (100310) in more detail, highlighting all the detected lines (see Appendix~\ref{app:lineid}). This eYSC has the brightest \pah emission in this sample, and its spectrum has several features connected with the early, embedded stages of star formation, suggesting a young evolutionary stage. We detect multiple H$_2$ transitions in the $S$, $Q$ and $O$ series, which in addition with the strong PAH emission, indicate a bright PDR still associated with this eYSC. There is a rise in the continuum level towards redder wavelengths which is associated with hot dust emission and typically observed in embedded star-forming regions \citep[e.g.,][]{woods_2011MNRAS.411.1597W}. 100310's spectrum also shows a $^{12}$CO$_2$ ice feature at 4.27~$\mu$m, and gas-phase CO transitions in the $R$ and $P$ branches, both originating from the cold, natal molecular cloud. Finally, we also detect some dust emission from deuterated hydrocarbon particles at $4.65~\mu$m for 100310 (and also several other of the brightest eYSCs in this sample). This deuterated emission presents as broadened wings around the Pf$\beta$ recombination line. Recently, this dust emission feature was also observed in M51 \citep{draine_2025ApJ...984L..42D} and in the Orion Bar \citep{peeters_2024A&A...685A..74P}. 

The other eYSC spectra in Figure~\ref{fig:source_spectra_all} show strong variations in the intensity of the different emission lines and spectral features observed. Among them, we also find signs of more evolved eYSCs. Figure~\ref{fig:source_spectra_all} shows some eYSCs (e.g., 200256) with distinct CO absorption features (i.e., $^{12}$CO bandheads, with the first CO overtone at $\sim2.3~\mu$m) and a change in the continuum shape between 1.5 and 2~$\mu$m. These are typical of older clusters hosting more evolved stars such as RSGs (\citealt{martins_2012A&A...547A..17M}). 

We will discuss the spectral properties of these eYSCs, as well as their immediate ISM, in the following subsections. %As was already mentioned, the present work is a presentation paper of the FEAST Cycle~2 NIRSpec/MOS observations of NGC\,628 (GO3503), and focuses on a small star-forming complex to test our data reduction pipeline, and to showcase the spectral analysis that is feasible with JWST in nearby galaxies. As such, the present work offers a taste of the properties of the young star clusters still emerging from their natal cloud, and thus actively driving feedback, in a single region in NGC\,628, but this will be expanded to the full galaxy observations in forthcoming publications.  

\subsection{Spectral properties of emerging young star clusters and their associated \hii regions}
\label{sec:eysc_spectra}

\begin{figure*}
    \centering
    \includegraphics[width=\linewidth]{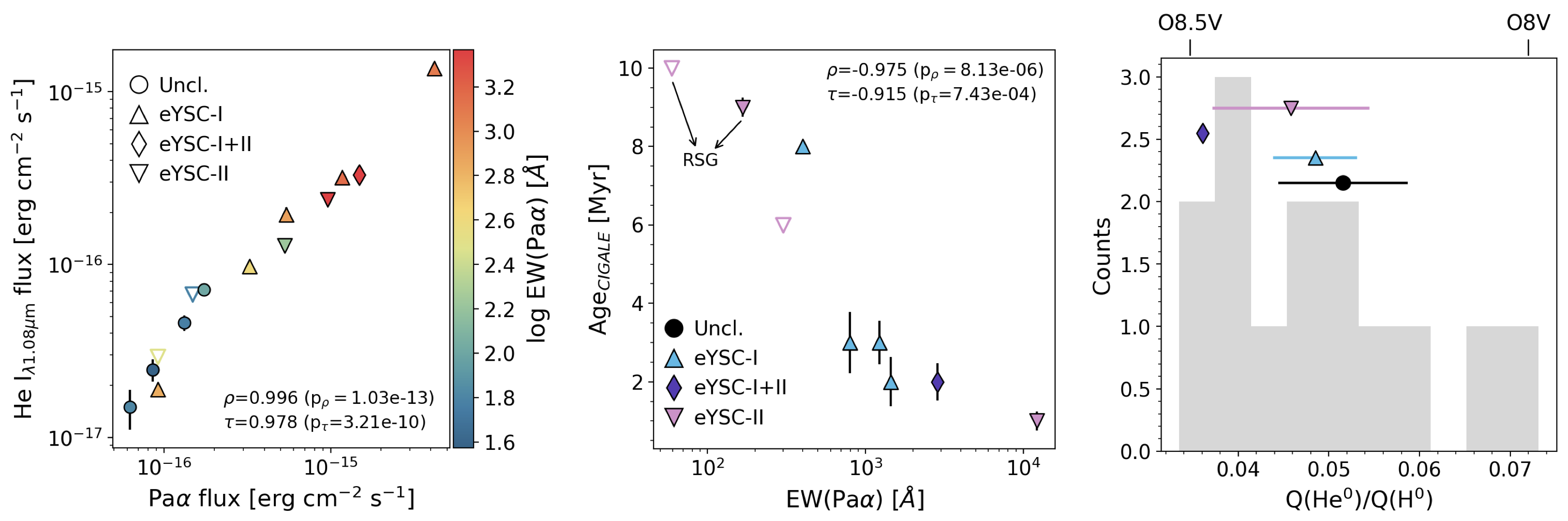}
    \caption{\textit{Left}: He~I ($\lambda=1.08\mu\mathrm{m}$) against Pa$\alpha$ flux for the eYSCs, colored by the equivalent width of Pa$\alpha$, EW(Pa$\alpha$). Upright triangles denote the eYSC-I, downward triangles the eYSC-II, diamonds the blended eYSC-I+II, and the circles the unclassified sources. The unfilled markers denote eYSCs for which no extinction correction was applied (see Sec.~\ref{sec:ext_corr}). The Spearman's $\rho$ and Kendall's $\tau$ statistics are shown in the bottom right corner (with associated p-values). \textit{Middle:} eYSC age estimates against EW(Pa$\alpha$). The age is estimated from SED fitting using \texttt{CIGALE} (see text; \citealt{pedrini_2025ApJ...992...96P}). The unclassified sources do not have available SED fits. Additionally, there is no age estimate for 1 eYSC (100291). eYSCs with RSG signatures in their spectra are highlighted (see text). Same marker type as left panel, and color scheme of Figure~\ref{fig:source_spectra_all}. \textit{Right}: Distribution of the Q(He$^0$)/Q(H$^0$) ratio for the eYSCs (shaded gray). Star spectral types likely powering the observed Q(He$^0$)/Q(H$^0$) are indicated at the top \citep{draine_2011piim.book.....D}. The markers (same as middle panel) represent the median Q(He$^0$)/Q(H$^0$) values for the different eYSCs, with the interquartile spread shown as the horizontal lines.}
    \label{fig:heipaa_age_qratio}
\end{figure*}

The feedback from massive stars directly affects the ionized gas of \hii regions. We focus here on characterizing the massive stellar content of the eYSCs, using the ionized gas as a tracer. The Pa$\alpha$ ($1.87~\mu$m) and He~I ($1.08~\mu$m) lines are the strongest H and He recombination lines in the $1-5~\mu$m wavelength range (under the assumptions of case B recombination and no reddening). Figure~\ref{fig:heipaa_age_qratio} (left panel) shows a very strong linear correlation between the Pa$\alpha$ and He~I fluxes for the eYSCs, regardless of eYSC class. Therefore, it is likely that similar populations of young, massive and hot stars in the eYSCs are producing the observed emission. The strong linear relation also suggests that the observed He~I emission is dominated by thermal rather than collisional excitation, since Pa$\alpha$ is not typically collisionally excited. Additionally, the equivalent width of the Pa$\alpha$ line, EW(Pa$\alpha$), can be related to the age of the clusters, with young ages typically associated with high EW(Pa$\alpha$).

As was already mentioned, the eYSC catalog that served as a foundation for these NIRSpec observations is built from the JWST imaging obtained by the FEAST survey in Cycle 1 (GO1783). All the available photometric bands between $0.2-5~\mu$m were used to analyze the spectral energy distribution (SED) of the FEAST eYSCs \citep[][S. T. Linden et al. \textit{in prep}]{pedrini_2025ApJ...992...96P}. The SEDs were fit with \texttt{CIGALE} \citep{boquien_2019A&A...622A.103B} to constrain cluster properties such as age \citep[see][for more details on the SED fitting]{pedrini_2025ApJ...992...96P}. In the middle panel of Figure~\ref{fig:heipaa_age_qratio} we show a comparison between the \texttt{CIGALE} age estimates and the spectroscopic Pa$\alpha$ equivalent width, for the eYSCs where both estimates are available. We see a good agreement between the two: eYSCs with the highest EW(Pa$\alpha$), and brightest H and He emission (left panel), are found by \texttt{CIGALE} to be the youngest ($<3$~Myr). This is consistent with expectations of earlier evolutionary stages producing more intense ionizing radiation. Additionally, all the eYSCs that have RSG signatures in their spectra (i.e., CO bandheads at $2.3\sim2.4~\mu$m, see Figure~\ref{fig:source_spectra_all}) all have low EW(Pa$\alpha$) and are found to be older ($>9$~Myr) from the SED fitting, consistent with expectations. Despite the good agreement between these photometric and spectroscopy age estimates, neither should be treated as absolute age indicators, since stochasticity can cause degeneracy. Recently, \cite{pedrini_2025ApJ...992...96P} have shown that \texttt{CIGALE} has limitations when constraining ages for NIR-detected eYSCs. The authors also show that for clusters with masses lower than $5,000~\mathrm{M}_\odot$, where stochastic initial mass function (IMF) sampling is significant, similar values of EW(Pa$\alpha$) can correspond to a wide range of cluster ages. %This could indeed be the case for source 200309, for example, as this eYSC-II has the highest EW(Pa$\alpha$) and lowest fitted age ($\sim$1~Myr) in this sample. The \texttt{CIGALE} mass estimate for this eYSC is $<1,000~\mathrm{M}_\odot$, and thus we expect stochasticity to have a significant effect \citep{pedrini_2025ApJ...992...96P}. 

The $1-5~\mu$m wavelength range of the MOS observations also allows us to further probe the ionized gas and stellar properties. Given that He~I emission requires more energetic photons ($E>24.6$~eV) compared to hydrogen recombination lines ($E>13.6$~eV), the ratio between the two can inform us on the average spectral type of the stars within the eYSCs that are driving the observed ionization. For that purpose, we measure the ionizing photon flux capable of exciting H, Q(H$^0$), and He, Q(He$^0$), following \cite{pasquali_2011AJ....141..132P}:

\begin{equation}
    \mathrm{Q(H^0)} [\mathrm{s^{-1}}] = 7.3\times10^{11} \, \left(8.5\times L(\mathrm{Pa\alpha}) [\mathrm{erg\,s^{-1}}] \right)
    \label{eqn:qh0}
\end{equation}
\begin{equation}
    \mathrm{Q(He^0)} [\mathrm{s^{-1}}] = 1.0\times10^{12} \, L(\mathrm{He~I}) [\mathrm{erg\,s^{-1}}]
\end{equation}

\noindent The Pa$\alpha$ and He~I luminosities, $L(\mathrm{Pa\alpha})$ and $L(\mathrm{He~I})$ respectively, are calculated from the fluxes through \mbox{$L=4\pi D^2 F$}, where $D$ is the distance to the galaxy (in cm), and $F$ is the relevant flux. We convert the original $L(\mathrm{H\alpha})$ in \cite{pasquali_2011AJ....141..132P} to $L(\mathrm{Pa\alpha})$ assuming a conversion factor of 8.5 \citep[assuming $T_e=10^4$~K and $n_e=10^3~\mathrm{cm}^{-3}$;][]{draine_2011piim.book.....D}. The distribution of Q(He$^0$)/Q(H$^0$) ratios for the eYSCs studied here is shown in Figure~\ref{fig:heipaa_age_qratio} (right panel). We measure a median of Q(He$^0$)/Q(H$^0$)$\simeq0.048\,(\pm0.008)$, indicating that O8.5V to O8V stars are dominating the emission from the eYSCs \citep{draine_2011piim.book.....D}. It is however likely that we are underestimating the true amount of ionizing photons, since a non-negligible amount of these could be quickly absorbed by dust within the \hii regions before it is able to ionize the gas \citep{choi_2020ApJ...902...54C,dellabruna_2021A&A...650A.103D}. %We will further discuss this topic in Sec.~\ref{sec:discussion}. 

\begin{figure}
    \centering
    \includegraphics[width=0.9\linewidth]{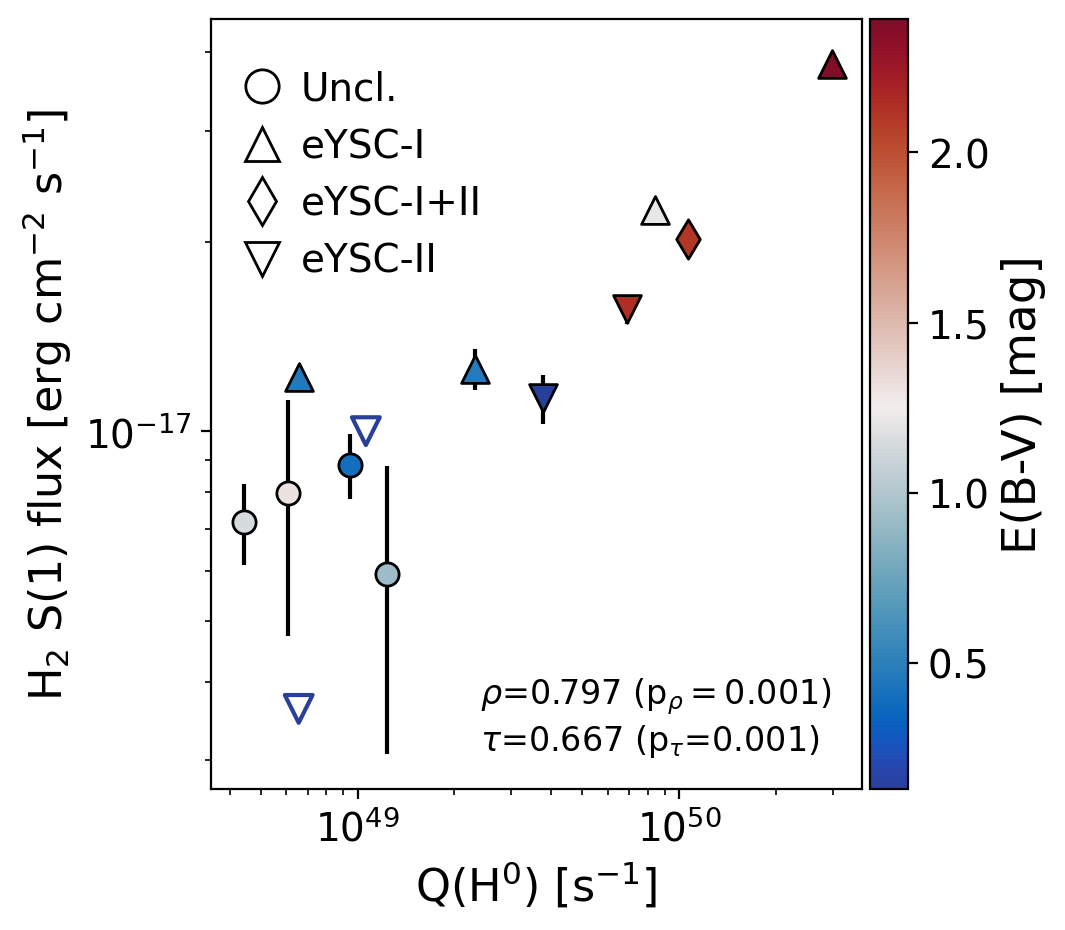}
    \caption{H$_2~1-0~S(1)$ line flux against hydrogen-ionizing photon flux, Q(H$^0$), colored by measured extinction, $E(B-V)$. Same marker type as Figure~\ref{fig:heipaa_age_qratio}, with unfilled markers denoting the eYSCs for which no extinction was applied ($E(B-V)=0$, see Sec.~\ref{sec:ext_corr}). The Spearman's $\rho$ and Kendall's $\tau$ statistics are shown in the bottom right corner (with associated p-values). We omit 1 eYSC from this plot, as its H$_2~ S(1)$ line is too faint for a reliable flux measurement.}
    \label{fig:h2_qh0}
\end{figure}

\begin{figure}
    \centering
    \includegraphics[width=0.9\linewidth]{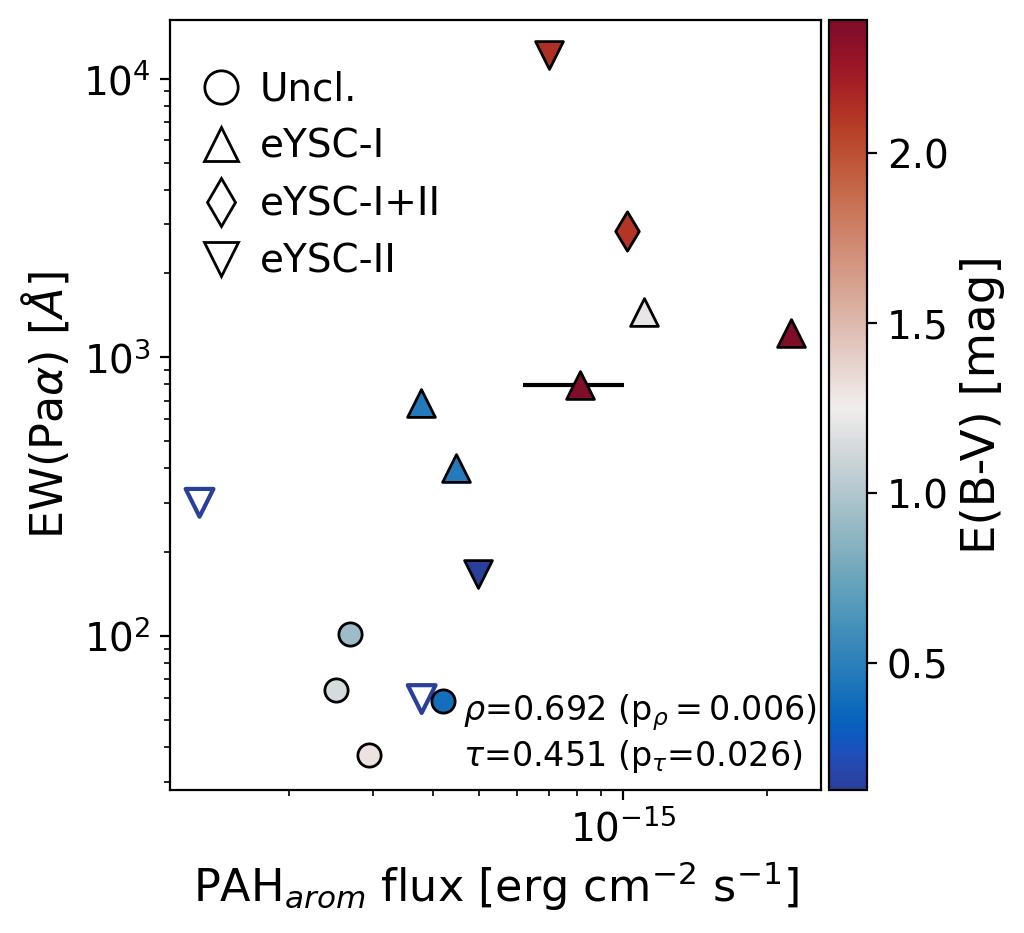}
    \caption{Equivalent width of Pa$\alpha$ line, EW(Pa$\alpha$), against the PAH$_{arom}$ flux at 3.3~$\mu\mathrm{m}$, colored by measured extinction, $E(B-V)$. Same marker type as Figure~\ref{fig:heipaa_age_qratio}, with unfilled markers denoting the eYSCs for which no extinction was applied ($E(B-V)=0$, see Sec.~\ref{sec:ext_corr}). The Spearman's $\rho$ and Kendall's $\tau$ statistics are shown in the bottom right corner (with associated p-values).}
    \label{fig:ew_pah}
\end{figure}

\subsection{Spectral properties of PDRs}
\label{sec:pdr_eysc}

At the early stages of the cluster emergence sequence, eYSCs are still associated with a compact PDR, which is particularly bright in warm H$_2$ and \pah emission \citep[e.g.,] []{peeters_2024A&A...685A..74P}. When looking at the amount of warm molecular gas present in the eYSCs (Figure~\ref{fig:h2_qh0}), we can see that more energetic eYSCs are also associated with brighter H$_2$ emission. Additionally, we observe higher \mbox{$E(B-V)$} values for these eYSCs, suggesting that they are more embedded in their natal cloud. These objects with high Q(H$^0$) and H$_2$ fluxes are indeed likely to be the youngest in the sample, since high Q(H$^0$) values (i.e., high Pa$\alpha$ fluxes) are associated with younger ages (see Figure~\ref{fig:heipaa_age_qratio}, Eqn.~\ref{eqn:qh0}).

As was already mentioned, the NIR PAH emission is particularly susceptible to the radiation field \citep{schroetter_2024A&A...685A..78S,lai_2025arXiv250904662L}, and there has been evidence of PDR morphology changing as eYSCs emerge from their natal cloud \citep{pedrini_2024ApJ...971...32P}. As can be seen from Figure~\ref{fig:ew_pah}, the age of eYSCs (as quantified by the EW(Pa$\alpha$)) and the \pah emission arising from their PDR have a moderate positive correlation (Spearman's rank and Kendall correlation coefficients of $\rho=0.69$ and $\tau=0.45$, respectively). In other words, it seems that as eYSCs evolve from their younger, more embedded state, to becoming more exposed (lower $E(B-V)$ values), the \pah emission decreases. Together with the similar trend seen with H$_2$ (see Figure~\ref{fig:h2_qh0}), this indicates that PDRs are indeed affected by the cluster emergence process.

\begin{figure}
    \centering
    \includegraphics[width=0.9\linewidth]{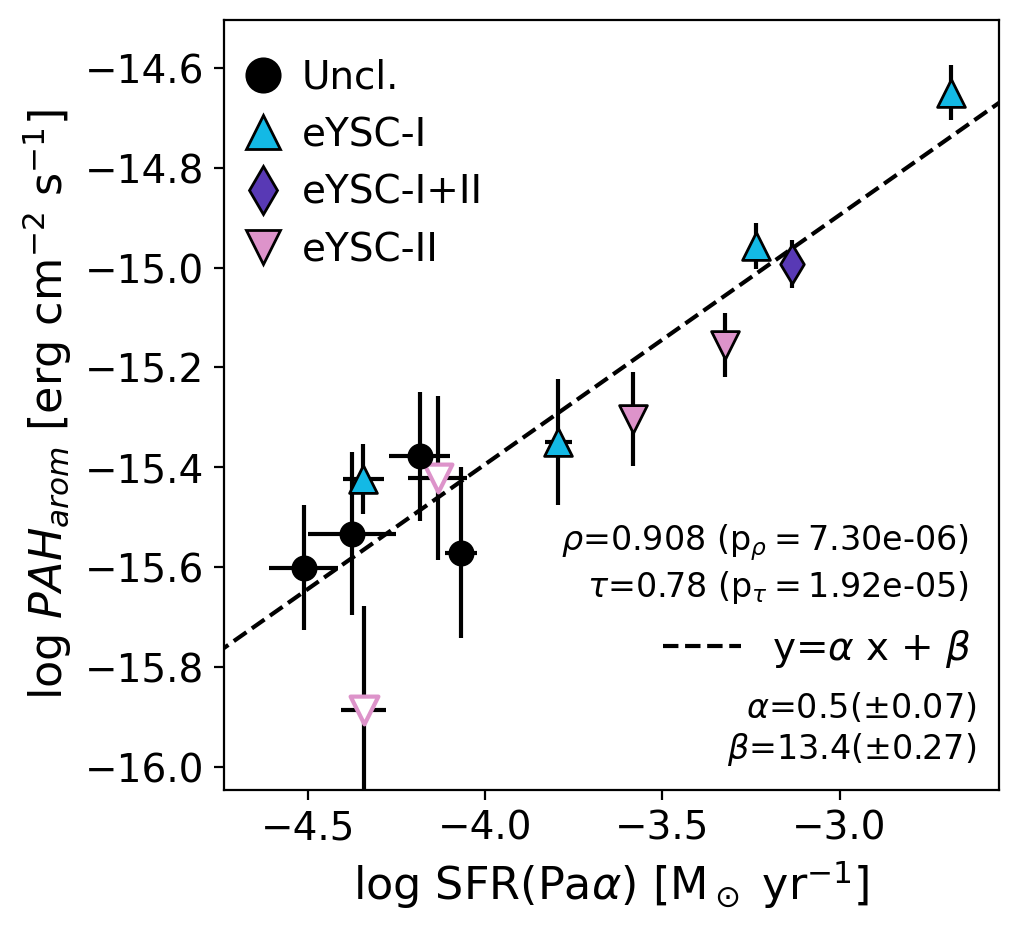}
    \caption{Log-log relation between the PAH$_{arom}$ flux at 3.3~$\mu\mathrm{m}$ and the star formation rate measured from Pa$\alpha$, SFR(Pa$\alpha$). Same marker type as Figure~\ref{fig:heipaa_age_qratio}, with unfilled markers denoting the eYSCs for which no extinction was applied ($E(B-V)=0$, see Sec.~\ref{sec:ext_corr}). The best fit relation is determined with Bayesian linear regression and is shown as the dashed black line (slope and intercept quoted in the bottom right corner). We also quote the Spearman's $\rho$ and Kendall's $\tau$ statistics (with associated p-values). We omit 1 eYSC from this plot due to high relative uncertainties.}
    \label{fig:pah_sfr}
\end{figure}

Due to the spatial correlation between \hii regions and PDRs, \pah emission is often used as a star formation rate (SFR) tracer across cosmic time \citep[e.g.,] []{lai_2020ApJ...905...55L,calzetti_2024ApJ...971..118C,gregg_2024ApJ...971..115G,gregg_2025arXiv251106481G}. Figure~\ref{fig:pah_sfr} shows the relation between the 3.3~$\mu$m PAH emission and the SFR measured from the Pa$\alpha$ flux of the eYSCs in this work. We convert the Pa$\alpha$ luminosity, L(Pa$\alpha$), to a SFR following \cite{calzetti_2024ApJ...971..118C}:

\begin{equation}
    \mathrm{SFR(Pa\alpha)} [\mathrm{M_\odot \, yr^{-1}}] = 4.26\times10^{-11} \, L(\mathrm{Pa\alpha}) 
\end{equation}

\noindent As can be seen from Figure~\ref{fig:pah_sfr}, we retrieve a strong positive correlation between PAH and SFR ($\rho=0.91$ and $\tau=0.78$). We fit the two quantities with \texttt{LINMIX}\footnote{\url{https://github.com/jmeyers314/linmix}}, a Bayesian linear regression algorithm that accounts for uncertainties on both variables. We find a sub-linear relation with slope $0.5\pm0.08$ and intercept $-13.4\pm0.28$, shown as a dashed black line in Figure~\ref{fig:pah_sfr}. We retrieve a similar sub-linear relation (slope of $0.5\pm0.34$ and intercept $-14.2\pm1.2$) and correlation coefficients ($\rho=0.89$ and $\tau=0.72$) between SFR(Pa$\alpha$) and the aliphatic component of the NIR PAH feature (PAH$_{aliph}$ at 3.4~$\mu$m) albeit with larger uncertainties given the faintness of this component. Recently, \cite{gregg_2024ApJ...971..115G} also found a sub-linear 3.3~$\mu$m PAH-SFR relation using JWST/NIRcam imaging, which holds against various continuum subtraction methodologies \citep[see also][]{gregg_2025arXiv251106481G}. This sub-linearity hints at PAH destruction in environments with more intense ionization, but that generally PAH emission can still be robustly used as a SFR tracer. Nevertheless, even though the PAH-SFR relation observed here is consistent with the literature, this work only focuses on a small spiral arm region of NGC\,628, and thus the galaxy-wide environmental variations are not represented. Future work will further populate Figure~\ref{fig:pah_sfr} with the full MOS spectroscopic dataset that covers a large portion of NGC\,628's disc (see Figure~\ref{fig:fullngc628}), and thus provide a more conclusive analysis on the PAH-SFR relation.

\begin{figure*}
    \centering
    \includegraphics[width=\linewidth]{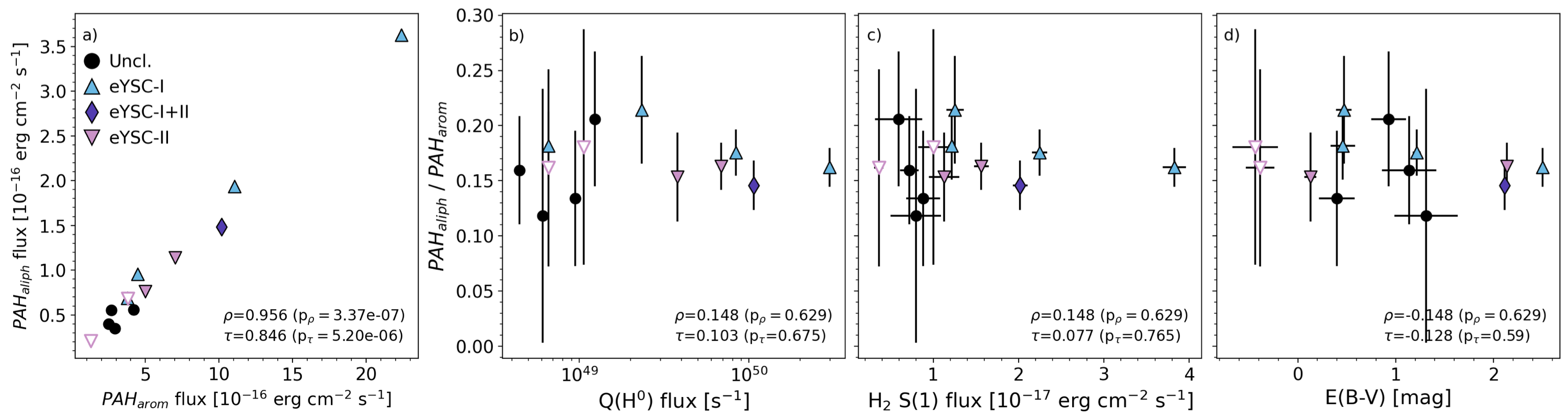}
    \caption{Properties of the NIR PAH feature at 3.3~$\mu$m for the eYSCs studied in this work. \textbf{a)} Aliphatic component, PAH$_{aliph}$ (at 3.4~$\mu$m), against the aromatic component, PAH$_{arom}$ (at 3.3~$\mu$m). Same marker type as Figure~\ref{fig:heipaa_age_qratio}, with eYSC-I in blue, eYSC-II in pink, blended eYSC-I+II in purple, and unclassified sources in black. \textbf{b)} Ratio between the aliphatic and aromatic components, PAH$_{aliph}$/PAH$_{arom}$, against measured extinction $E(B-V)$. \textbf{c)} PAH$_{aliph}$/PAH$_{arom}$ against H$2~S(1)$ flux. \textbf{d)} PAH$_{aliph}$/PAH$_{arom}$ against hydrogen-ionizing photon flux, Q(H$^0$). For all subplots, the Spearman's $\rho$ and Kendall's $\tau$ statistics are also shown in the bottom right corner (with associated p-values). }
    \label{fig:pahratio}

\end{figure*}

The multiple components of the NIR PAH feature also carry information about the degree of UV exposure experienced by the PAHs \citep[e.g.,] []{schroetter_2024A&A...685A..78S}. The aromatic component, PAH$_{arom}$ at 3.3~$\mu$m, is attributed to aromatic C-H bonds while the aliphatic component, PAH$_{aliph}$ at 3.4~$\mu$m, corresponds to aliphatic C-H stretch emission \citep{allamandola_1989ApJS...71..733A}. Aliphatic C-H bonds are more prone to photodissociation when exposed to UV radiation than the aromatic C-H bonds \citep{marciniak_2021A&A...652A..42M}. Thus, a small ratio of PAH$_{aliph}$/PAH$_{arom}$ can indicate UV-processed PAHs, whilst high PAH$_{aliph}$/PAH$_{arom}$ suggests the PAHs are more shielded from UV radiation \citep{peeters_2024A&A...685A..74P,lai_2023ApJ...957L..26L}. In the Orion Bar PDR, \cite{schroetter_2024A&A...685A..78S} find PAH$_{aliph}$/PAH$_{arom}\simeq0.04$ for irradiated, UV-exposed regions, and PAH$_{aliph}$/PAH$_{arom}\simeq0.1$ for shielded PAHs \citep[see also][]{lai_2023ApJ...957L..26L}. Figure~\ref{fig:pahratio} shows the measured PAH$_{arom}$ and PAH$_{aliph}$ fluxes for the eYSCs studied in this work, as well as their ratio (PAH$_{aliph}$/PAH$_{arom}$) as a function of hydrogen-ionizing photon flux, Q(H$^0$), H$_2$ flux, and extinction. The PAH components display a strong positive correlation with each other ($\rho=0.96$ and $\tau=0.85$), with a median ratio of PAH$_{aliph}$/PAH$_{arom}=0.162\pm0.027$, suggesting a minimal degree of UV exposure. Additionally, there is no apparent trend of PAH$_{aliph}$/PAH$_{arom}$ with $E(B-V)$, H$_2$ or Q(H$^0$). We discuss this lack of trend further in Sec.~\ref{sec:discussion}.

\subsection{Spectral properties of the diffuse ISM}
\label{sec:ism_spectra}

\begin{figure*}
    \centering
    \includegraphics[width=\linewidth]{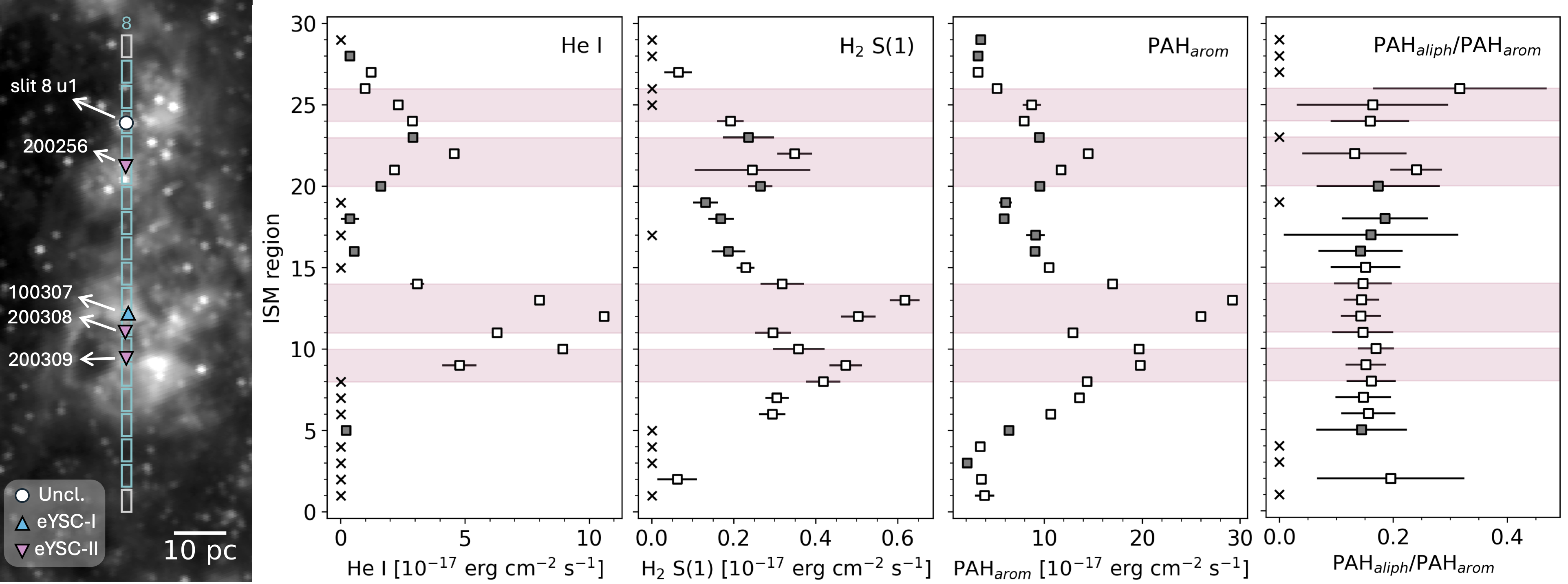}
    \caption{Spectral characterization of the ISM in a star-forming complex, which is possible due to the long slit configurations in the FEAST NIRSpec/MOS observations. The leftmost plot contains a vertical view of the MOS slit 8, with the NIRCam/F335M in the background, which is continuum-subtracted to reveal the \pah emission (same as Figure~\ref{fig:fullngc628}). The 4 plots on the right showcase the measured fluxes of He~I, H$_2~S(1)$, PAH$_{arom}$, and the PAH$_{aliph}$/PAH$_{arom}$ ratio (respectively) along the length of the slit. This slit is subdivided into 29 different apertures, and ISM spectra are extracted for each one (i.e., ISM regions on the y-axis, see text). For each of these 4 panels, the eYSC positions within the slit (see Sec.~\ref{sec:source_spectra_extr}) are highlighted in pink. The squares represent the respective flux measurements, color-coded as white where extinction correction was applied (from Br$\alpha$/Pa$\beta$, see text), while gray squares represent the ISM regions for which no extinction was reliably measured (and thus these fluxes are lower limits). Black crosses denote non-measurements as well as flux measurements with high relative uncertainties ($>100\%$).}
    \label{fig:ism_spectra}
\end{figure*}

%- The primary goal of this NIRSpec/MOS survey is to characterize eYSCs, and thus the depth and sensitivity of the observations are attuned to this.

Star-forming regions are incredibly complex structures, truly showcasing the multi-phase nature of the ISM. This is particularly evident in Galactic star-forming regions such as the Orion Bar, where we can pick out the \hii region, its ionization front, multiple dissociation fronts of the PDR, and the surrounding, more diffuse ISM \citep[see][]{habart_2024A&A...685A..73H}. Outside of the Local Group we cannot reach this level of precision yet; however, JWST's sensitivity and resolution does allow us to study the variations of the more diffuse ISM within these star-forming complexes.

In this Section, we investigate the spectral properties of the diffuse ISM as a function of distance from the ionizing eYSCs (see Sec.~\ref{sec:ism_spectra_extr}). Figure~\ref{fig:ism_spectra} depicts the local variations of ionized gas (traced by He~I), molecular gas (H$_2$) and PAH emission along the longest MOS slit studied here (slit 8, shown on the leftmost panel of the figure). In the Figure, we highlight the positions of the eYSCs within the slit for guidance (shaded regions). As expected, the He~I emission is highly localized, and concentrated where eYSCs are located. On the other hand, the molecular gas and PAH emission are more spatially extended, with both peaking at the locations of eYSCs, and decreasing as a function of increasing distance to these ionizing sources. This more extended emission is expected, since less energetic photons can escape the \hii regions and drive excitation further away \citep[e.g.,] []{calapa_2014ApJ...784..130C,sandstrom_2023ApJ...944L...8S,pedrini_2024ApJ...971...32P}. It follows that these escaping photons have a higher chance of ionizing H$_2$ and PAH closer to the \hii regions than further away, giving rise to this declining trend. We observe no discernible trend between PAH$_{aliph}$/PAH$_{arom}$ ratio and distance from the ionizing eYSCs, with values of $\sim0.15-0.2$ across the entire slit (similar to Sec.~\ref{sec:pdr_eysc}).      

The primary goal of the FEAST NIRSpec/MOS observations is to characterize the spectral properties of eYSCs (and associated PDRs). As such, the depth and sensitivity of the observations are attuned for this, rather than the fainter emission from the neighboring ISM. Still, in Figure~\ref{fig:ism_spectra} we show that we can measure emission line fluxes for majority of the diffuse ISM within the chosen slit, despite the small apertures (see Sec.~\ref{sec:ism_spectra_extr}). We could not reliably measure $E(B-V)$ for a few of these ISM regions, and only lower limits of the emission line fluxes were retrieved (gray boxes in Figure~\ref{fig:ism_spectra}). This is not due to non-detections of either Br$\alpha$ or Pa$\beta$ (i.e., the lines used for the extinction correction), as Pa$\beta$ (the fainter of the two) is measured with at least S/N$=3$ for all ISM regions. Instead, this is likely an effect of the assumed temperature and electron density values when deriving the intrinsic line ratio, which here we keep fixed at $T_e=10^4$~K and $n_e=10^3~\mathrm{cm}^{-3}$ (i.e., typical \hii region assumption).  Thus, as we step away from \hii regions, the observed line ratios become unphysical in comparison with the intrinsic ratio, implying very little extinction. In any case, the relative variation in the spectral properties shown in Figure~\ref{fig:pahratio} is not likely to change much with different temperature and density assumptions.

\subsection{Diagnosing the stellar feedback}
\label{sec:stellar_feedback}

\begin{figure}
    \centering
    \includegraphics[width=0.9\linewidth]{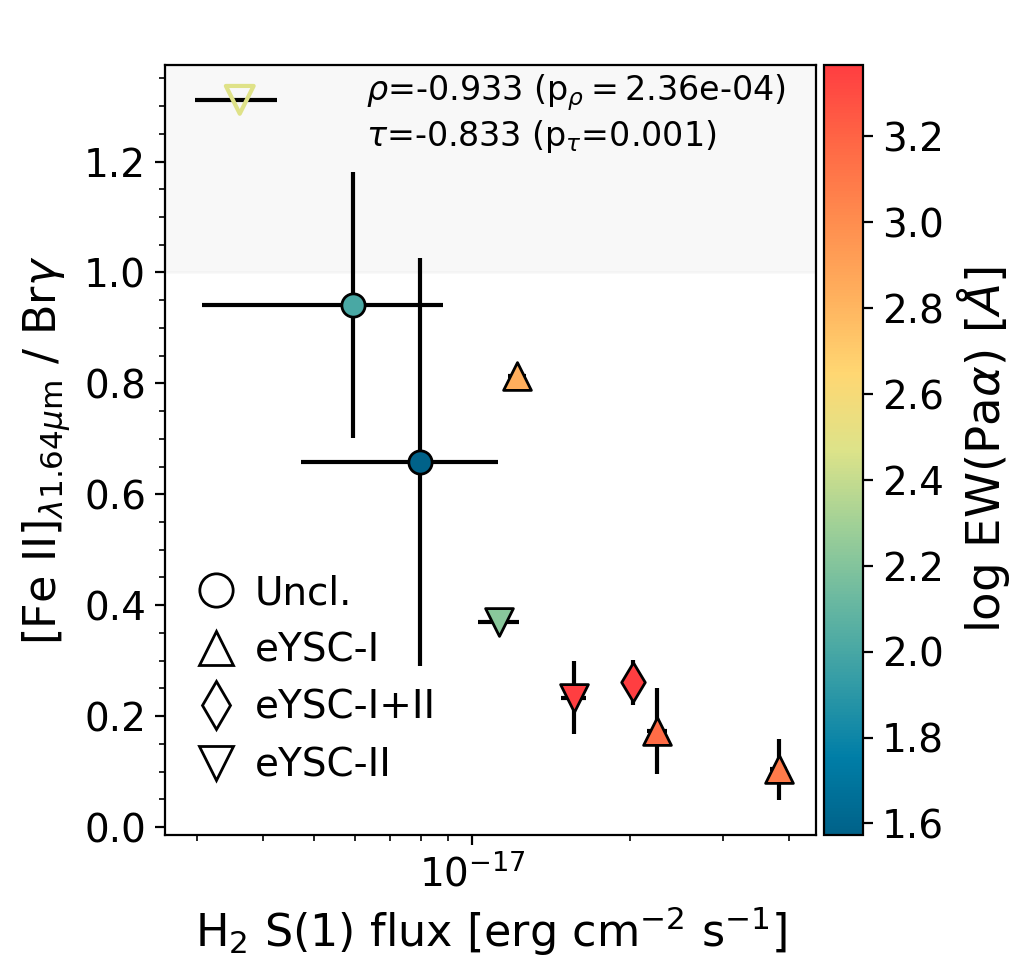}
    \caption{[Fe\,II]/Br$\gamma$ ratio (shock tracer) against H$_2~1-0~S(1)$ line flux, colored by Pa$\alpha$ equivalent width, EW(Pa$\alpha$). Same marker type as Figure~\ref{fig:heipaa_age_qratio}, with unfilled markers denoting the eYSCs for which no extinction was applied (see Sec.~\ref{sec:ext_corr}). The Spearman's $\rho$ and Kendall's $\tau$ statistics are shown in the top right corner (with associated p-values). We omit eYSCs with relative uncertainties on measured fluxes higher than 60\%.}
    \label{fig:FeII/brg_h2}
\end{figure}

The previous sections have highlighted that the eYSC sample studied here is at an early stage of the cluster emergence process, since their emission is dominated by young, hot massive stars (Sec.~\ref{sec:eysc_spectra}) and are still associated with a bright PDR within the natal cloud (Sec.~\ref{sec:pdr_eysc}). Taking all this into consideration, as well as the young age estimates for the eYSCs (median of $\sim3~$Myr from the SED fitting, see Figure~\ref{fig:heipaa_age_qratio}), it is probable that pre-SNe feedback is dominant. To test this, we use NIR line diagnostics to constrain the nature of the active stellar feedback processes. The [Fe\,II]$_{\lambda1.64\mu\mathrm{m}}$/Br$\gamma$ ratio can be used to distinguish between regimes dominated by SNe-driven shocks ([Fe\,II]/Br$\gamma \gg 1$) and photoionization ([Fe\,II]/Br$\gamma < 1$) \citep{alonso-herrero_1997ApJ...482..747A,cresci_2010A&A...520A..82C}. As can be seen from Figure~\ref{fig:FeII/brg_h2}, the observed values of [Fe\,II]/Br$\gamma$ for the eYSCs are small ($\lesssim1$), consistent with the photoionization regime. There is a strong negative correlation between [Fe\,II]/Br$\gamma$ and H$_2$, with smaller values of [Fe\,II]/Br$\gamma$ corresponding to brighter H$_2$ emission ($\rho=-0.93$ and $\tau=-0.83$, considering only eYSCs with less than 60\% relative uncertainty on the [Fe\,II]/Br$\gamma$ ratio). The correlation remains strongly negative considering instead all eYSCs with [Fe\,II] detections (11/14 eYSCs with $\rho=-0.8$, $\mathrm{p}_\rho=0.003$ and $\tau=-0.7$, $\mathrm{p}_\tau=0.003$), and also those with relative uncertainties less than 30\% ($\rho=-0.89$ with $\mathrm{p}_\rho=0.02$ and $\tau=-0.73$ with $\mathrm{p}_\tau=0.05$). Furthermore, eYSCs at the earliest stages of the emergence sequence (i.e., high EW(Pa$\alpha$)) display small values of [Fe\,II]/Br$\gamma$ and consequently larger H$_2$ fluxes, whilst older, more evolved eYSCs show less associated H$_2$ and increasing [Fe\,II]/Br$\gamma$ ratios. Altogether, this indicates that the eYSCs studied here are driving stellar feedback that is mostly radiative in nature. 

To corroborate this, we also measure the \mbox{H$_2~1-0~S(1)/2-1~S(1)_{\lambda2.248\mu\mathrm{m}}$} ratio, which can help constrain the excitation mechanism behind the observed H$_2$ emission. Radiative fluorescent excitation of H$_2$ results in \mbox{H$_2~1-0~S(1)/2-1~S(1)\simeq2$}, whilst \mbox{H$_2~1-0~S(1)/2-1~S(1)\simeq10$} is typically seen for thermal excitation due to shocks \citep{davies_2003ApJ...597..907D,martin_hernandez_2008A&A...489..229M}. Following the methodology outlined in Sec.~\ref{sec:line_fluxes}, we measured the H$_2~2-1~S(1)$ flux (at $2.248\mu\mathrm{m}$) for 11 eYSCs (3 non-detections), and retrieved an average of \mbox{H$_2~1-0~S(1)/2-1~S(1)\simeq2.04\pm0.76$}, with values in the $1-3$ range. The average H$_2$ ratio (as well as the observed range) comfortably falls in the radiative FUV heating regime. Indeed, all eYSCs with \mbox{[Fe\,II]/Br$\gamma<0.5$} display a median H$_2$ ratio of $1.98\pm0.2$. From the values of \mbox{[Fe\,II]/Br$\gamma\lesssim1$} and \mbox{H$_2~1-0~S(1)/2-1~S(1)\simeq2$}, we can conclude that the observed eYSCs are driving a FUV radiation field that is capable of exciting the molecular gas in the PDR, with little contribution from shocks \citep{black_1987ApJ...322..412B}. 

\begin{figure*}
    \centering
    \includegraphics[width=0.83\linewidth]{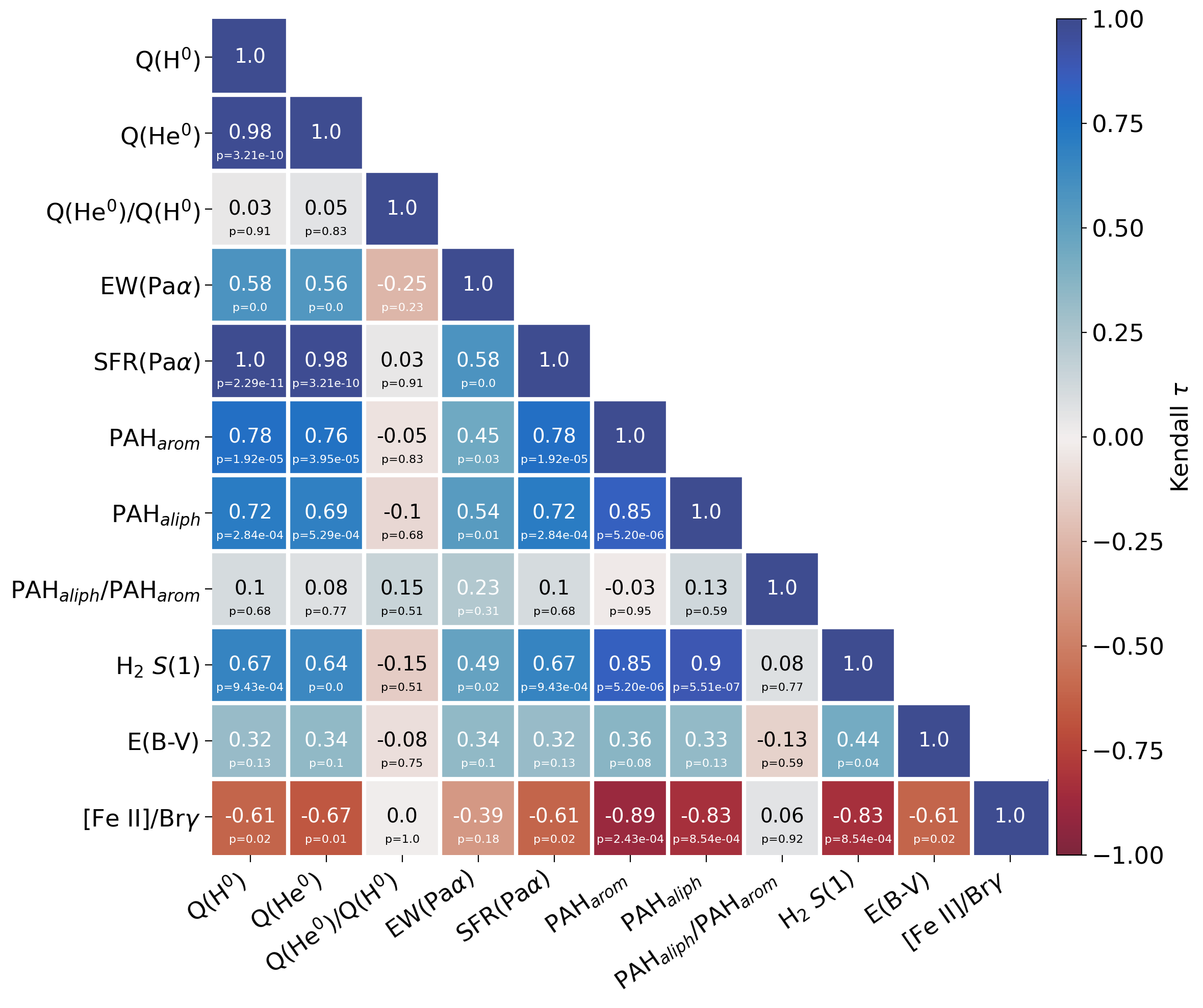}
    \caption{Kendall $\tau$ correlation ladder for all spectral properties of eYSCs measured in this work. These include: hydrogen-ionizing photon flux, Q(H$^0$); helium-ionizing photon flux, Q(He$^0$); the ratio of the former two, Q(He$^0$)/Q(H$^0$); Pa$\alpha$ equivalent width, EW(Pa$\alpha$); star formation rate derived from Pa$\alpha$, SFR(Pa$\alpha$); aromatic component of PAH feature at 3.3~$\mu$m, PAH$_{arom}$; aliphatic component of PAH feature at 3.4~$\mu$m, PAH$_{aliph}$; the ratio of the former two, PAH$_{aliph}$/PAH$_{arom}$; H$_2$ $S(1)$ line flux; measured extinction, $E(B-V)$; and the shock-tracer [Fe\,II]/Br$\gamma$. For each box, the Kendall $\tau$ value is written, with the associated p-value below, and color-coded following the colorbar.}
    \label{fig:corr_matrix}
\end{figure*}

\section{Discussion and future work}
\label{sec:discussion}

This work highlights the power of JWST/NIRSpec in characterizing the spectral properties of emerging young star clusters, and surrounding ISM, in galaxies beyond the Local Group. With spectroscopy in the $1-5~\mu$m range, we can single out the characteristics of the ionization and stellar content of the eYSCs through He and H recombination lines, the PDR properties with H$_2$ and \pah emission, and probe the active stellar feedback mechanisms via line diagnostics. This paper introduces the FEAST Cycle 2 NIRSpec/MOS observations of NGC\,628 (GO3503), which target a significant sample of eYSCs across this star-forming spiral galaxy. The initial results presented here focus on a small subsample of eYSCs (14 out of 176) within a $\sim0.5\times0.5~\mathrm{kpc}^2$ spiral arm region of the galaxy, meant to test our data reduction strategy and showcase the quality of the observations and planned scientific analysis. Forthcoming publications will include the full spectroscopic dataset, which spans the disc of NGC\,628, thus allowing for an investigation into environmental variations in spectral properties (e.g., spiral arms versus inter-arm). In this Section, we discuss our results in context with the literature, and highlight areas for further work. Figure~\ref{fig:corr_matrix} displays a Kendall's $\tau$ correlation ladder of the spectral properties analyzed in this work, which serves as a summary of our findings.  

\subsection{The stellar properties of eYSCs}

We find that the majority of the eYSCs studied here are at an early and energetic evolutionary phase, consistent with expectations from the photometry \citep[e.g.,] []{pedrini_2024ApJ...971...32P}. The eYSCs are associated with bright Pa$\alpha$ ($1.87~\mu$m) and He~I ($1.08~\mu$m) emission, implying a stellar population capable of producing energetic photons. The sources with the highest H and He fluxes also tend to be the youngest, as estimated by the equivalent width of Pa$\alpha$ (Figure~\ref{fig:heipaa_age_qratio}). Additionally, we characterize the stellar content dominating the emission from the eYSCs by calculating the ionizing photon fluxes capable of exciting H and He, Q(H$^0$) and Q(He$^0$) respectively. We find a median of \mbox{Q(He$^0$)/Q(H$^0$)$=0.048(\pm0.008)$}, which is consistent with O8.5V to O8V stars \citep{draine_2011piim.book.....D}. %As mentioned in Sec.~\ref{sec:eysc_spectra}, it is very likely that the ionizing photon fluxes measured here are underestimated \citep[e.g.,] []{choi_2020ApJ...902...54C,dellabruna_2021A&A...650A.103D}. Given that helium-ionizing photons are inherently more energetic, these should be absorbed by dust at higher rates than the hydrogen-ionizing photons. Thus, both Q(H$^0$) and Q(He$^0$) are lower limits, but Q(He$^0$) is affected more. This implies higher Q(He$^0$)/Q(H$^0$) ratios than those measured here, and consequently a more energetic, massive and young O star type. Regardless, 
It is clear that the eYSCs studied here host hot, young massive stars which are capable of ionizing He, and thus are at an early, energetic evolutionary phase. We see no correlation between the type of radiation field powered by the eYSCs (traced by Q(He$^0$)/Q(H$^0$), i.e., the dominating spectral type of the stellar population) with any other quantity (see Figure~\ref{fig:corr_matrix}). However, the small subset of eYSCs studied in this work, in addition with the small range of Q(He$^0$)/Q(H$^0$) values observed (see right panel of Figure~\ref{fig:heipaa_age_qratio}), can very easily explain this lack of correlation. We will revisit this topic for the full spectroscopic dataset.  

The eYSC ages estimated both from the photometry \citep[][S. T. Linden et al. \textit{in prep}]{pedrini_2025ApJ...992...96P} and from the equivalent width of the Pa$\alpha$ emission line (middle panel of Figure~\ref{fig:heipaa_age_qratio}) corroborate the conclusion that these clusters are at an early energetic phase. The eYSC subsample in this work is found to have a median age of $\sim3$~Myr from the \texttt{CIGALE} SED fits, and a median EW(Pa$\alpha$) value of $\sim6400~\AA$. Furthermore, the youngest, most ionizing eYSCs typically display higher $E(B-V)$ values, and are associated with more molecular gas (i.e., H$_2~1-0~S(1)$ at $2.12~\mu$m) and \pah emission (see Figures~\ref{fig:h2_qh0}, \ref{fig:ew_pah} and \ref{fig:corr_matrix}). Altogether, this indicates that the eYSCs have a compact PDR, and are still embedded in their natal molecular cloud. 

In particular, the eYSC with the brightest PAH emission in this subsample (100310) also has several spectral features associated with a deeply embedded phase (e.g., $^{12}$CO$_2$ ice absorption and CO gas-phase transitions; see Figure~\ref{fig:spectra_zoom}). The CO$_2$ ice absorption feature has also been recently detected in bright star-forming regions in M51 \citep{draine_2025ApJ...984L..42D}. The asymmetric profile of this feature is likely a result of scattering effects, which affects primarily the short-wavelength wing \citep[e.g.,][]{brunken_2024A&A...692A.163B}. Interestingly, we do not detect H$_2$O ice (at $3~\mu$m), which could have implications in the ice formation pathways of this region \citep[e.g.,][]{smith_2025NatAs...9..883S}, but further analysis is required. We will further investigate this embedded source with follow-up ALMA Cycle 12 observations of the cold molecular gas (traced by $^{12}$CO) in this region. These will allow us to characterize the natal molecular cloud at the same physical resolution as the JWST imaging ($\sim0.2^{\prime\prime}$ or $10$~pc), directly connecting the eYSC properties to the parent cloud.

On the other hand, the eYSCs that show CO bandhead absorption features in their spectra ($\sim2.3~\mu$m, see Figure~\ref{fig:source_spectra_all}), which are typical signatures of more evolved RSG stars \citep{martins_2012A&A...547A..17M}, are found to have lower values of EW(Pa$\alpha$), and \texttt{CIGALE} fitted ages \mbox{$>9$~Myr}, consistent with expectations. Future work with the full spectroscopic dataset will investigate these objects further, and compare the age estimates from the \texttt{CIGALE} SED fitting to more tailored age measurements for star clusters with RSGs, which use the equivalent width of the first CO overtone at $2.3~\mu$m compared to predictions from \texttt{starburst99} \citep{starburst99_2014ApJS..212...14L} for different star types \citep[e.g.,] []{martins_2012A&A...547A..17M}.

\subsection{The connection between eYSCs and PDRs}

As can be seen from Figure~\ref{fig:corr_matrix}, we find a strong correlation between the ionized gas (both H and He), H$_2$, and PAH emission, indicating a tight connection between the eYSCs and their PDRs. As clusters move along the evolutionary sequence and begin emerging from their natal cloud (i.e., towards higher ages and lower $E(B-V)$ values), there is a decrease in both PAH and H$_2$ emission (see Figures~\ref{fig:h2_qh0} and \ref{fig:ew_pah}). This could be interpreted as an evolution of the morphology of PDRs associated with the eYSCs as these move along the cluster emergence sequence. \cite{pedrini_2024ApJ...971...32P} also find that \mbox{\pah} emission decreases as a function of eYSC age \citep[see also][]{linden_2024ApJ...974L..27L}, and that the morphology of PDRs varies depending on the evolutionary stage of the eYSCs: from compact to open for more evolved, exposed clusters. The strong correlation between H$_2$ and PAH emission (shown in Figure~\ref{fig:corr_matrix}) suggests that both species are present in PDRs, and likely being excited by the same radiation field that is powered by the eYSCs. This is further supported by the trends seen in the diffuse ISM spectra (Figure~\ref{fig:ism_spectra}), where PAH and H$_2$ emission display similar variations as a function of distance from the ionizing eYSCs.

Additionally, the tight connection between the ionized gas and the PAH emission originating from the PDRs observed here and in other works \citep[e.g.,][]{gregg_2025arXiv251106481G} is often exploited, with the latter often being used as a SFR indicator across cosmic time \citep[e.g.,][see also Sec.~\ref{sec:pdr_eysc}]{calzetti_2024ApJ...971..118C}. Similar to \cite{gregg_2024ApJ...971..115G}, we observe a sub-linear PAH-SFR relation (slope of $0.5\pm0.08$ in log-space), with a small scatter of $\sim0.09$~dex (defined as the mean orthogonal distance between the best-fit relation and the data points). However, as was already discussed in Sec.~\ref{sec:pdr_eysc}, the PAH-SFR relation found here may not be representative of the full galaxy, and should not be used for more general purposes. Forthcoming work will re-derive this relation for the full NIRSpec/MOS dataset.

In this work, we decompose the PAH emission in the spectra into different components: PAH$_{arom}$ at $3.3~\mu$m, and PAH$_{aliph}$ at $3.4~\mu$m. These different components arise from different C-H bonds \citep{allamandola_1989ApJS...71..733A}. The aliphatic bonds that result in the PAH$_{aliph}$ emission are more easily photodissociated than the aromatic component \citep{marciniak_2021A&A...652A..42M}. Consequently, the PAH$_{aliph}$/PAH$_{arom}$ ratio has been used to trace the degree of UV exposure experienced by the PAH dust grains \citep{schroetter_2024A&A...685A..78S}. We find a very strong correlation between the two PAH components (see Figure~\ref{fig:pahratio}), with a median PAH$_{aliph}$/PAH$_{arom}$ ratio of $0.16\pm0.03$. This suggests that the PAHs are shielded, and thus experiencing a minimal degree of UV exposure that is capable of photodissociating the aliphatic component. It could be that we simply do not achieve the resolution necessary to differentiate the ``shielded" versus ``irradiated" phases of PDRs as is possible to do in the Orion Bar \citep{schroetter_2024A&A...685A..78S,chown_2024A&A...685A..75C}, and are instead observing the outer envelope of PAHs in PDRs, whereas the irradiated phase is located closer to the \hii region. For II~Zw~40, a galaxy at a similar distance as NGC\,628 (10~Mpc) albeit with lower metallicity, \cite{lai_2025arXiv250904662L} also found \mbox{PAH$_{aliph}$/PAH$_{arom}\simeq0.1$} \citep[see also][]{lai_2023ApJ...957L..26L}, and no variation with ionization field hardness (probed by mid-IR line diagnostics). As can be seen in Figure~\ref{fig:corr_matrix}, we find no correlation between PAH$_{aliph}$/PAH$_{arom}$ and any other quantity studied in this work, including the eYSC ionizing output and H$_2$ emission (see Figure~\ref{fig:pahratio}). We also do not observe any variations in PAH$_{aliph}$/PAH$_{arom}$ as a function of distance from eYSCs (Figure~\ref{fig:ism_spectra}). Although both PAH components individually scale with H$_2$ emission, the lack of trend between PAH$_{aliph}$/PAH$_{arom}$ and H$_2$ could suggest that the PAH composition is resilient to the local variations that drive different molecular excitation levels, at least at the physical scales that we are able to probe here.   

\subsection{Stellar feedback}

Finally, in this work we also investigate the nature of the stellar feedback produced by the eYSCs. We distinguish between a feedback regime dominated by SNe-driven shocks from a photoionization-dominated regime using two independent NIR line diagnostics: [Fe\,II]/Br$\gamma$ \citep{cresci_2010A&A...520A..82C}, and \mbox{H$_2~1-0~S(1)/2-1~S(1)_{\lambda2.25\mu\mathrm{m}}$} \citep{martin_hernandez_2008A&A...489..229M}. In this sample of eYSCs, we find sub-unity values of [Fe\,II]/Br$\gamma$ and \mbox{H$_2~1-0~S(1)/2-1~S(1)\simeq2$}, which are both consistent with a photoionization-dominated regime, with little contribution from shocks (i.e., SNe). This again highlights that pre-SNe feedback is a major component of the star formation cycle, at least when He-ionizing, massive O stars are present, in line with results from the literature \citep{barnes_2021MNRAS.508.5362B,mcleod_2021MNRAS.508.5425M,menon_2023MNRAS.521.5160M,pathak_2025ApJ...982..140P}. In this framework, SNe explode in an already processed ISM and thus can efficiently create large cavities (i.e., bubbles) and disperse molecular clouds by clearing the leftover natal gas \citep{chevance_2022MNRAS.509..272C,grudic_2022MNRAS.512..216G,andersson_2024A&A...681A..28A}.

%- Future work, explore the dust emission arising from deuterated hydrocarbon particles at $4.65~\mu$m to further explore the deuteration of hydrocarbon particles such as PAHs, and any correlations with hardness and intensity of the radiation field \citep{draine_2025ApJ...984L..42D}.

\section{Summary and conclusions}
\label{sec:conclusion}

Here, we present a first look at the FEAST Cycle 2 JWST/NIRSpec multiplex spectroscopy observations (G03503) of emerging young star clusters in a nearby galaxy (NGC\,628, $\sim10$~Mpc). In this presentation paper, we characterize the spectral properties of an initial sample of 14 eYSCs, as well as the photodissociation regions, and the diffuse ISM within a $\sim0.5\times0.5~\mathrm{kpc}^2$ region. Our findings further support that the FEAST eYSC catalog consists of star clusters at early evolutionary stages, still partly embedded in the natal cloud and actively driving stellar feedback. We use near-infrared line diagnostics to constrain the dominant stellar feedback mechanisms. We summarize our results as follows:

\begin{itemize}
    \item We detect numerous He and H recombination lines tracing the \hii regions powered by the eYSC, as well as multiple H$_2$ transitions and bright \pah emission originating from the PDRs still associated with these young clusters. Furthermore, we observe $^{12}$CO bandhead absorption features for some clusters, indicating the presence of more evolved red supergiant stars. On the other hand, we observe an eYSC with CO$_2$ ice absorption in its spectra, as well as CO gas-phase features, implying that this eYSC is still embedded in the cold, dense, natal molecular cloud. Altogether, these findings highlight the power of JWST in unveiling the very early stages of the star formation cycle in galaxies outside the Local Group.

    \item The eYSCs are at an early and energetic evolutionary phase, with hot, young massive stars with high ionizing outputs. The measured ionizing photon fluxes are consistent with O8.5V-O8V stars dominating the emission from the eYSCs. %It is however very likely that the ionizing photon fluxes are underestimated, and that the average spectral type is even more energetic than that measured.

    \item The observed equivalent widths of the Pa$\alpha$ line suggest that these eYSCs are indeed young, and are consistent with the ages derived from photometric SED fits \citep[median $\sim3$~Myr;][S. T. Linden et al. \textit{in prep}]{pedrini_2025ApJ...992...96P}. The eYSCs that have spectral signatures consistent with more evolved stars (i.e., red supergiants) have smaller equivalent widths, and have age estimates $>9$~Myr, as expected. 

    \item The ionized gas is found to be highly correlated with both the molecular gas emission (i.e., H$_2$) and the $3.3~\mu$m PAH emission. As the star clusters age and begin to emerge from their natal cloud, both the H$_2$ and PAH emission decrease. This indicates a tight connection between eYSCs and their PDR, with PDR morphology evolving as the clusters emerge from their natal cloud \citep[see also][]{pedrini_2024ApJ...971...32P}.

    \item Utilizing the spatial direction of the NIRSpec/MOS slits, we found localized ionized gas tracing the \hii regions, and more spatially extended H$_2$ and $3.3~\mu$m PAH emission. The H$_2$ and PAH emission display similar profiles as a function of distance from the ionizing eYSCs and are highly correlated, suggesting that the same radiation field is likely exciting both species.

    \item We find a sub-linear relationship between $3.3~\mu$m PAH emission and the SFR from Pa$\alpha$, with a power-law exponent of $0.5\pm0.08$ and a strong positive correlation ($\rho=0.91$ and $\tau=0.78$). This strong correlation further supports the use of $3.3~\mu$m PAH as a SFR indicator \citep[e.g.,] []{kennicutt_2009ApJ...703.1672K}, but the observed sub-linearity hints at PAH destruction in more intense radiation field environments \citep[see][]{gregg_2024ApJ...971..115G,gregg_2025arXiv251106481G}. %However, the small subsample of eYSCs studied here, which are restricted to a small region within NGC\,628, make this PAH-SFR relation purely empirical and not intended for general use.  

    \item The FEAST NIRSpec/MOS observations resolve the individual aromatic and aliphatic components, centered at $3.3$ and $3.4~\mu$m respectively, of the broad NIR PAH feature. Since the aliphatic component is more prone to photodissociation \citep{marciniak_2021A&A...652A..42M}, the aliphatic/aromatic ratio can be used as a probe of the UV exposure experienced by the PAHs \citep{schroetter_2024A&A...685A..78S}. We find \mbox{PAH$_{aliph}$/PAH$_{arom}=0.16\pm0.03$}, suggesting that generally the PAHs in this region are being shielded from UV exposure capable of photodissociating aliphatic C-H bonds \citep{lai_2025arXiv250904662L}. However, it could be that we are only observing the outer shielded envelope of shielded PAHs, and cannot probe the irradiated phase at these physical scales.

    \item We investigate the active stellar feedback mechanisms via emission line diagnostics. We retrieve \mbox{[Fe\,II]$_{\lambda1.64\mu\mathrm{m}}$/Br$\gamma\lesssim1$} and \mbox{H$_2~1-0~S(1)/2-1~S(1)=2.04\pm0.76$}, which are both consistent with a photoionization-dominated regime, with little contributions from SNe-driven shocks \citep{martin_hernandez_2008A&A...489..229M,cresci_2010A&A...520A..82C}.
    
\end{itemize}

Here, we showcase the type of spectral analysis that is now feasible in nearby galaxies thanks to JWST's capabilities. The richness of spectral features observed, at a level of detection that is unprecedented for a galaxy at this distance, allow us to trace key components of the star formation cycle. For the first time, we can reveal the spectral properties of the emerging star clusters, their PDR, and probe the active stellar feedback mechanisms outside of the Local Group. This paper presents an initial analysis of a portion of the FEAST NIRSpec/MOS observations of NGC\,628, demonstrating the potential of the data. Future work will deliver the first spectral characterization of a statistically significant sample of eYSCs across the disk of a galaxy. With the full spectroscopic dataset, we will investigate any trends with galactic environment, and compare the findings with photometry-based studies \citep[e.g.,][S. T. Linden et al. \textit{in prep}]{pedrini_2025ApJ...992...96P}. In particular, we will take advantage of the multiple H$_2$ transitions observed (see Figure~\ref{fig:spectra_zoom}) to build H$_2$ excitation ladders, which will allow us constrain the column density and temperature of the molecular gas of PDRs \citep[e.g.,][]{martin_hernandez_2008A&A...489..229M,peeters_2024A&A...685A..74P,hunt_2025ApJ...993...84H}. 

This work highlights the great potential of the MOS mode of NIRSpec for star formation studies in nearby galaxies. This paper constitutes a proof-of-concept, and shows the significant payout in both sensitivity and resolution when sampling hundreds of embedded star clusters at the same time, for a relatively small commitment of telescope time.

%% Please use the acknowledgment and contribution environments. This will 
%% be anonomyized when the ``anonymous" style option is used. 
\begin{acknowledgments}

This work is based on observations made with the NASA/ESA/CSA James Webb Space Telescope, which is operated by the Association of Universities for Research in Astronomy, Inc., under NASA contract NAS 5-03127. The data were obtained from the Mikulski Archive for Space Telescopes (MAST) at the Space Telescope Science Institute. The observations presented in this work are associated with program \#3503. Support for program \#3503 was provided by NASA through a grant from the Space Telescope Science Institute, 
which is operated by the Association of Universities for Research in Astronomy, Inc., under NASA contract NAS 5-03127. HFV and AA acknowledge support from the Swedish National Space Agency (SNSA) through the grant 2023-00260. AA and AP acknowledge support from SNSA through grant 2021-00108. ADC acknowledges the support from a Royal Society University Research Fellowship (URF/R1/19160 and URF/R/241028). KG is supported by the Australian Research Council through the Discovery Early Career Researcher Award (DECRA) Fellowship (project number DE220100766) funded by the Australian Government. 

HFV especially thanks Anna de Graaff and Alberto Saldaña-Lopez for useful discussions which improved this manuscript.
\end{acknowledgments}

\begin{contribution}
%%This section gives authors the space to recognize author contributions. The text inside this environment is NOT counted towards the total word quanta. At a minimum, manuscripts are expected to include this text:

All authors contributed equally to this manuscript.

%% But authors are expected to provide more specific details, e.g., 
%%
%%SC was responsible for writing and submitting the manuscript.
%%WWM came up with the initial research concept and edited the manuscript.
%%OTS obtained the funding and edited the manuscript.
%%EBF provided the formal analysis and validation. He also edited the manuscript.
%%GEH Supervised the undergraduates, wrote the software and administers the project github and Zenodo repositories.
%%
%% Authors can use the Contributor Role Taxonomy (CRediT) at
%% https://credit.niso.org
%% for ideas on how write a good statement tailored to their needs.

\end{contribution}

%% To help institutions obtain information on the effectiveness of their 
%% telescopes the AAS Journals has created a group of keywords for telescope 
%% facilities.
%
%% Following the acknowledgments section, use the following syntax and the
%% \facility{} or \facilities{} macros to list the keywords of facilities used 
%% in the research for the paper.  Each keyword is check against the master 
%% list during copy editing.  Individual instruments can be provided in 
%% parentheses, after the keyword, but they are not verified.
\facilities{JWST(NIRSpec/MOS, NIRCam)}
%% Similar to \facility{}, there is the optional \software command to allow 
%% authors a place to specify which programs were used during the creation of 
%% the manuscript. Authors should list each code and include either a
%% citation or url to the code inside ()s when available.
\software{astropy \citep{astropy_2013A&A...558A..33A,astropy_2018AJ....156..123A,astropy_2022ApJ...935..167A}, numpy, sci-kit, jwst-pipeline, pyneb \citep{pyneb_2015A&A...573A..42L}, dust\_extinction \citep[][see also \citealt{gordon_2009,fitzpatrick_2019,decleir_gordon_2022,gordon2023ApJ...950...86G}]{gordonpackage2024JOSS....9.7023G}, msafit \citep{degraaff_msafit_2024A&A...684A..87D}, linmix 
}

%% Appendix material should be preceded with a single \appendix command.
%% There should be a \section command for each appendix. Mark appendix
%% subsections with the same markup you use in the main body of the paper.
%%
%% Each Appendix (indicated with \section) will be lettered A, B, C, etc.
%% The equation counter will reset when it encounters the \appendix
%% command and will number appendix equations (A1), (A2), etc. The
%% Figure and Table counter will not reset.

\appendix
\counterwithin{figure}{section} 
\counterwithin{table}{section}

\section{Flux loss due to aperture size and shutter position: forward-modeling with \texttt{msafit}}
\label{app:msafit}

\begin{figure}
    \centering
    \includegraphics[width=\linewidth]{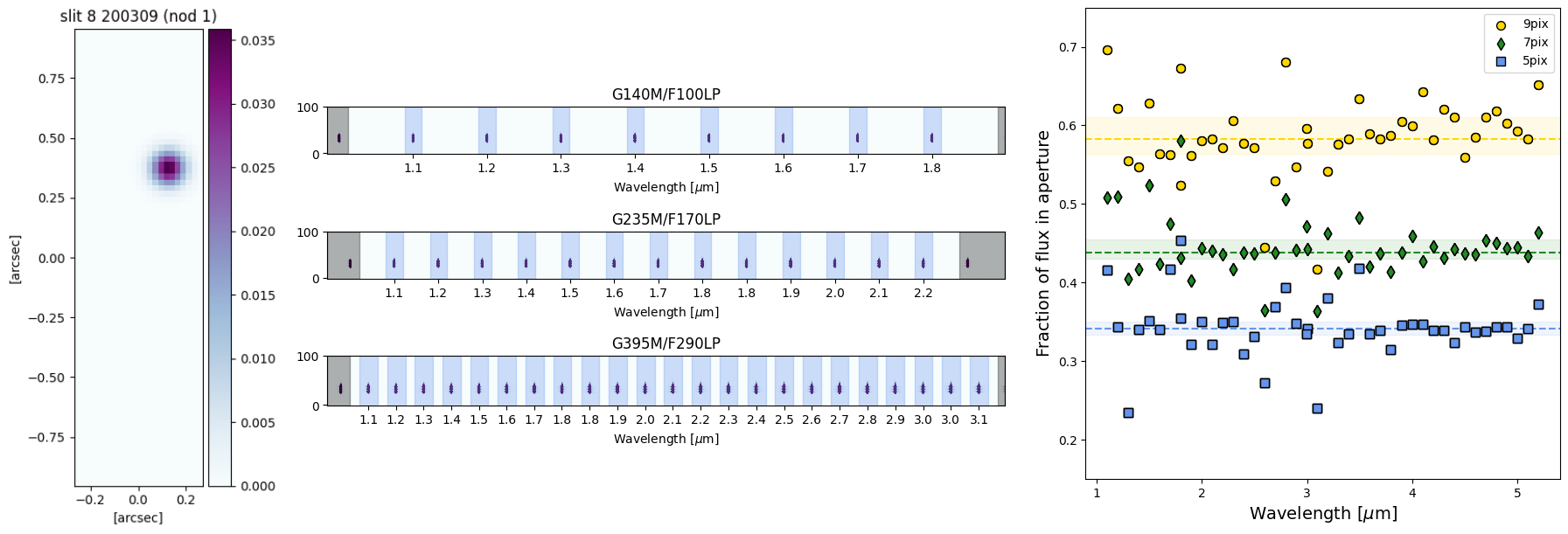}
    \caption{Overview of forward-modeling performed with \texttt{MSAFIT}. \textit{Left}: Gaussian model of source 200309 (slit 8) with a half-light radius of $0.1^{\prime\prime}$. The source's placement in the MSA shutter follows the observed position. \textit{Middle}: Model 2D spectra resulting from the \texttt{MSAFIT} PSF convolution and frame rectification/combination from the JWST data reduction pipeline (see main text). \textit{Right:} Fraction of flux contained in different sized apertures, relative to the observed ``emission line" flux shown in the middle subplot (see main text) for source 200309. The yellow circles represent the flux fractions for a 9~pix aperture at a given wavelength, the green diamonds for a 7~pix aperture, and the blue squares for a 5~pix aperture. These correspond to physical sizes of 44.9, 34.9 and 24.9~pc, respectively. Following the same color-scheme, the dashed lines depict the median flux fraction in the respective aperture across all modeled wavelength points, and the shaded region shows the interquartile range.}
    \label{fig:app_msafit_aperture}
\end{figure}

We complement the radiometric steps in the JWST data reduction pipeline, which tackle various types of flux loss along the optical path, with forward-modeling performed with \texttt{MSAFIT} \citep{degraaff_msafit_2024A&A...684A..87D}. \texttt{MSAFIT} allows us to model a given source's morphology and its placement within the slit, and thus retrieve an estimate of how much extended flux is lost due to the source's position in the MSA shutter. We model each source in this work with a Gaussian profile, a half-light radius of $0.1^{\prime\prime}$, and an intrinsic flux of 1 (see left subplot of Figure~\ref{fig:app_msafit_aperture}). The modeled source is given the same position as the observed source (both in the detector plane and in the slit), which is taken from the MSA metadata reference file generated during the observations. We do not know the true position of unclassified sources in their slit, since these sources are only identified in the spectroscopy and do not exist in the photometric eYSC catalogs from FEAST (see Sec.~\ref{sec:source_spectra_extr}), and thus are not included in the MSA reference files. We model these sources with the correct detector plane placement (since this is given by the slit), but place them in the center of the MSA shutter. 

Apart from recovering the extended source flux lost due to the positioning within the MSA slit, we also measure the amount of flux that is lost due to the extraction aperture size we use to build our source 1D spectra (see Sec.~\ref{sec:source_spectra_extr}). Following \cite{degraaff_msafit_2024A&A...684A..87D} and \cite{slob_2024ApJ...973..131S}, we use \texttt{MSAFIT} to generate a 3D model cube of each source along a wavelength grid that samples the entire wavelength range of our observations, in $0.1~\mu$m intervals. An ``emission line" with a total normalized intensity of 1 and velocity dispersion of 0 is placed at each wavelength point. For each disperser/filter combination, the model cube is convolved with synthetic NIRSpec point-spread function (PSF) models, which account for wavelength dependencies \citep[see][for further details]{degraaff_msafit_2024A&A...684A..87D}. The cube is then modeled onto the NIRSpec/MOS detector plane assuming a standard $1\times3$ shutter configuration. This results in 18 unrectified 2D frames for each source: 6 for each grating, where each nodding in the 3-nod dither pattern is modeled onto both MOS detectors (NRS1 and NRS2). We do not attempt to forward-model sources that are only observed in one dither in this work (101724+100264 and 200156), and thus apply no further flux loss corrections to these sources. The \texttt{MSAFIT} unrectified 2D frames are equivalent to the outputs from Stage 1 of the JWST data reduction pipeline (see Sec.~\ref{sec:reduction_pipeline}). To capture any further flux losses arising from the rectification and combination of individual frames by the pipeline, the individual \texttt{MSAFIT} 2D frames are passed through Stage 2 and Stage 3 of the JWST pipeline in the same methodology applied to the real data. In this process, the radiometric steps of the pipeline are turned off, since for this exercise we are only interested in retrieving an estimate of how much of the modeled ``emission lines" are captured within a given aperture size, and consequently, how much of the flux is lost outside the aperture. We do not attempt to modify any of the radiometric steps in the pipeline, as this is outside the scope of this work.

Figure~\ref{fig:app_msafit_aperture} (middle panel) shows the resulting rectified, combined 2D spectra of the \texttt{MSAFIT} forward-modeling for source 200309 (slit 8). These modeled 2D spectra have the same format as the real, observed 2D spectra, and thus we can apply the same aperture extraction to calculate how much flux is lost in the process. The rightmost panel of Figure~\ref{fig:app_msafit_aperture} displays the fraction of the total flux contained in a given aperture size for each emission line modeled (i.e., for each wavelength point in the grid). As can be seen from the Figure, the flux contained within any sized aperture remains constant across wavelength, and thus it is reasonable to assume a single aperture for the spectral extraction across all 3 gratings. This remains the case for all sources. For the particular source in Figure~\ref{fig:app_msafit_aperture} (200309), we adopt an aperture of 7~pix for its extraction (see Figure~\ref{fig:s2d}), and our forward-modeling with \texttt{MSAFIT} estimates a flux loss of $\sim56\%$ given the chosen aperture. We compensate for these flux losses through:

\begin{equation}
    F(\lambda)_{new} = (1 + \mathrm{f}_{corr}) \times F(\lambda), 
    \label{eqn:app_flux_loss}
\end{equation}

\noindent where $F(\lambda)$ is the (uncorrected) flux density for a wavelength $\lambda$, $\mathrm{f}_{corr}$ the flux correction estimated from our \texttt{MSAFIT} forward-modeling (normalized to 1), and $F(\lambda)_{new}$ the corrected flux density. Table~\ref{tab:app_flux_corr} holds the calculated flux losses due to source placement in the MSA shutter and aperture size, as well as the flux correction implemented in this paper for each source.

\begin{table}
    \centering
    \begin{tabular}{c|c c | c}
        Source & Slit loss & Aperture loss & Flux correction ($\mathrm{f}_{corr}$) \\
        \hline
        200309 & 0.003 & 0.562 &  0.565 \\ 
        100307 + 200308 & 0.001 & 0.331 &  0.332
 \\ 
        200256 & 0.005 & 0.391 & 0.396 \\ 
        slit8 u1 & -- & 0.559 & 0.559 \\ 
        200255 & 0.066 & 0.590 & 0.656 \\ 
        101724 + 100264 & -- & -- & -- \\ 
        100310 & 0.004 & 0.468 & 0.472 \\ 
        100304 & 0.013 & 0.609 & 0.622 \\ 
        100291 & 0.002 & 0.779 & 0.781 \\ 
        slit63 u1 & -- & 0.82 & 0.82 \\ 
        slit63 u2 & -- & 0.644 & 0.644 \\ 
        slit64 u1 & -- & 0.86 & 0.86 \\ 
        100265 & 0.007 & 0.758 & 0.765 \\ 
        200156 & -- & -- & -- \\ 
    \end{tabular}
    \caption{Fractional flux losses due to placement within the MSA shutter (Slit loss) and spectral extraction aperture size (Aperture loss) for each eYSC in this work. The sum of these two gives the flux correction ($\mathrm{f}_{corr}$) applied in this work. These are estimated with \texttt{MSAFIT} (see text).}
    \label{tab:app_flux_corr}
\end{table}

\subsection{Comparison with photometry}

\begin{figure}
    \centering
    \includegraphics[width=\linewidth]{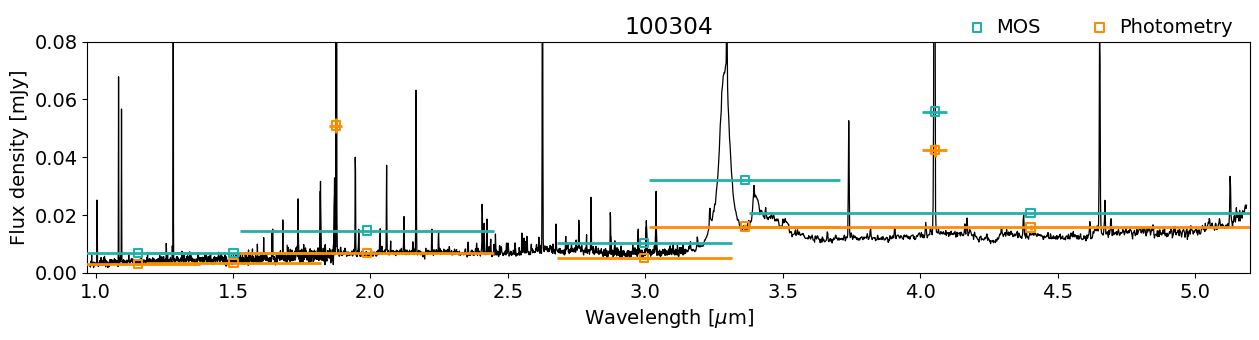}
    \caption{Comparison of flux density estimates of the MOS spectroscopy (turquoise) against photometry measurements from FEAST (orange; A. Adamo et al., \textit{in prep}; \citealt{gregg_2024ApJ...971..115G}) for a representative source (100304 in slit 62). The full MOS spectrum is shown in black. The square markers show the respective flux density estimates (with uncertainties as errorbars in the y-axis) for the relevant NIRCam bandpasses (wavelength range shown as errorbars in the x-axis).}
    \label{fig:phot_comp}
\end{figure}

The eYSCs analyzed in this work were first identified through NIRCam imaging (A. Adamo et al., \textit{in prep}; \citealt{gregg_2024ApJ...971..115G}), and thus have flux density estimates from aperture photometry for several NIRCam filters: F150W, F187N, F200W, F277W, F335M,
F405N, and F444W. To test our reduction pipeline for the MOS spectroscopy (including the aperture correction performed with \texttt{MSAFIT}), we convolve each source's spectrum with the JWST filter throughputs to estimate the MOS flux density in the corresponding NIRCam bands. Neither the photometry nor the spectroscopy is corrected for extinction here. Figure~\ref{fig:phot_comp} displays a representative example of this exercise. We retrieve consistent results between our spectroscopic flux densities and the photometry, with similar SED slopes, for all eYSCs. Any offsets naturally arise from the inherently different methodology in continuum-subtraction and aperture size and correction between the spectroscopy and the photometry.

\section{Line identification}
\label{app:lineid}

Given the level of detail present in the FEAST NIRSpec/MOS spectra (see Figure~\ref{fig:spectra_zoom}), the relatively long wavelength coverage ($1-5~\mu$m), and the large sample of spectra retrieved across the whole galaxy, the identification of features in the spectra had to be automated as much as possible. In this Appendix, we outline our first efforts in building an automated line identification code. 

The first step in this process is estimating the continuum level for each spectrum. Here, we use the observed wavelengths (i.e., not in rest frame), and correct the flux densities for extinction (see Sec.~\ref{sec:ext_corr}). The continuum estimation is done in an iterative fashion with the continuum fitting feature of \texttt{specutils} (\texttt{fit\_generic\_continuum}), which uses a high-order 1D polynomial. We first make an initial approximation of the continuum to identify any lines above or below a 10$\sigma$ level ($\sigma$ being the flux uncertainty or noise estimate at a given wavelength) from this initial estimate. This will also prevent any noise spikes from contaminating the continuum estimation. For the G395M spectra, the $3.2-3.6~\mu$m wavelength range is also not considered, as this corresponds to the very extended PAH feature at $3.3~\mu$m. These strong features are then masked out, and the final continuum fit is re-calculated for each spectra (each grating treated separately). We found that a reasonable single continuum fit was not possible for the eYSC spectra which present CO bandhead absorption features, and a rise in the continuum on the blue side of the spectra (i.e., spectra with RSG signatures, see Sec.~\ref{sec:results} and Figure~\ref{fig:source_spectra_all}). In these cases, we fit the continuum for the blue side of the spectrum (i.e., before the first CO bandhead at $\sim2.3~\mu$m) separately from the remaining part of the spectrum, and stitch both fits together.

Once the continuum is estimated and subtracted from the spectra, we begin the line identification process. Across each grating, we identify any peaks in either emission or absorption in the spectra which are above (below for absorption) a 5$\sigma$ level, and where the flux value of the identified peak is higher than the adjacent spectral pixels. The observed ``central" wavelength of each peak is then cross-matched with reference line lists. We use a reference list of observed NIR emission lines in the Orion Bar PDR from PDRs4All \citep{berne_2022PASP..134e4301B,peeters_2024A&A...685A..74P}, and a CO line list\footnote{\url{https://www.gemini.edu/observing/resources/near-ir-resources/spectroscopy/co-lines-and-band-heads}.} which includes the bandhead features (i.e., RSG signatures) as well as the fundamental CO $1-0$ lines at $4.5-5~\mu$m (see Sec.~\ref{sec:results}). We convert the rest central wavelengths of each line in the reference lists to the observed frame through the measured $z$ from the spectra ($z\simeq0.002$). For each identified peak, we cross-match its observed wavelength with the reference lists, and return all reference lines within $\pm0.002~\mu$m of this value. If an identified peak does not return any matches within the reference line lists, and if its flux density is above 10$\sigma$ for only 1 spectral pixel, we label this as a detector artifact, rather than a real spectral feature.

Some identified peaks in the spectrum have multiple matches within the reference line lists. We determine the most likely match, i.e., the line most likely causing the observed spectral feature, based on the reference versus observed central wavelength difference and line intensity. For each of the matches, we calculate the difference between the reference line central wavelength and the observed wavelength of the peak in the spectra (both in the observed frame). Where available, we weight this wavelength difference using the line intensity listed in the reference line list. For the PDRs4All observed line list, this line intensity is the integrated flux of the line from a Gaussian fit \citep[see Sec.~3.2 in][]{vandeputte_2024A&A...687A..86V}. This will ensure that we are not just getting the closest match in wavelength, which could be biased given the differences in spectral resolution between the two studies. We assume that between two reference lines, the one that is observed with a higher intensity in the Orion Bar PDR is more likely to drive the observed spectral feature in NGC\,628, which is much further away. In cases where no reference line intensities can be retrieved (such as the CO lines), we simply adopt the closest match in wavelength. To ensure no line identification is driven by errors in determining $z$, we ensure that any match must have a difference between the reference line central wavelength and observed peak central wavelength smaller than 5$\sigma_z$, where $\sigma_z$ is the standard deviation of the redshift measurements using several bright H recombination lines for the given spectra (see Sec.~\ref{sec:ext_corr}). 

We filter through this initial list of identified lines in search of any unexpected gaps in a series of transitions of the same spectral species. Particularly for H, the most abundant species in our spectra, we would not expect to detect the $24-4$ transition if $23-4$ is not detected. Therefore, any identified lines that appear after a gap in a series are removed from the list. This automated line identification code works pretty consistently for the spectra shown in this work, but further tests need to be conducted for the full dataset. Further work is also needed in flagging when multiple spectral features overlap, particularly when no intensity-weighting is available. %The continuum fitting performed in this exercise is also very simplified, as it is only meant to give a general fit across the whole spectra for the sole purpose of identifying lines. 

%We list the emission and absorption lines we are unable to identify in Appendix Table A1; these are due to warm pixels, fringe flat correction issues, undersampling, and stitching effects in overlap channels

%% For this sample we use BibTeX plus aasjournalv7.bst to generate the
%% the bibliography. The sample7.bib file was populated from ADS. To
%% get the citations to show in the compiled file do the following:
%%
%% pdflatex sample7.tex
%% bibtext sample7
%% pdflatex sample7.tex
%% pdflatex sample7.tex

\bibliography{main}{}
\bibliographystyle{aasjournalv7}

%% This command is needed to show the entire author+affiliation list when
%% the collaboration and author truncation commands are used.  It has to
%% go at the end of the manuscript.
%\allauthors

%% Include this line if you are using the \added, \replaced, \deleted
%% commands to see a summary list of all changes at the end of the article.
%\listofchanges

\end{document}